\documentclass[12pt,preprint]{aastex}

\usepackage{graphicx}
\usepackage[figuresright]{rotating}
\usepackage{lscape}

  \newcommand{\halpha}{$\rm{H}\alpha$}
  \newcommand{\hbeta}{$\rm{H}\beta$}
  \newcommand{\hgamma}{$\rm{H}\gamma$}
  \newcommand{\oi}{[\ion{O}{1}]}
  \newcommand{\oii}{[\ion{O}{2}]}
  \newcommand{\oiii}{[\ion{O}{3}]}
  \newcommand{\nii}{[\ion{N}{2}]}
  \newcommand{\sii}{[\ion{S}{2}]}
  \newcommand{\hii}{\ion{H}{2}}
  \newcommand{\feii}{\ion{Fe}{2}}
  \newcommand{\fevii}{[\ion{Fe}{7}]}
  \newcommand{\fex}{[\ion{Fe}{10}]}
  \newcommand{\hei}{\ion{He}{1}}
  \newcommand{\heii}{\ion{He}{2}}
  \newcommand{\kms}{km~s$^{-1}$}    
  \newcommand{\nd}{\nodata}         

\slugcomment{To appear in PASP.}

\shorttitle{Physical parameters of galactic nuclei from SBS}

\shortauthors{de Diego}

\begin{document}


\title{Physical Parameters and Classification of Eight Galactic
    Nuclei from the Second Byurakan Survey}

\author{Jos{\'e} A. de Diego}
\affil{Instituto de Astronom{\'\i}a\\ Universidad Nacional Aut{\'o}noma de
M{\'e}xico\\ Apartado Postal 70-264\\ M{\'e}xico D.F., 04510\\ Mexico}


\begin{abstract}
   Accurate spectral parameters estimates are essential to
   investigate the physical conditions in AGNs. Particularly,
   intrinsic reddening can be estimated from the \halpha/\hbeta\
   ratio of the narrow line components. The mass of the central
   black hole can be obtained from the stellar velocity dispersion
   in the bulge, which can be estimated from the \oiii\ $\lambda
   5007$ width, or by the relation between the AGN luminosity and
   the width of the broad \hbeta\ component. Two different
   luminosity values are important: one derived from the
   monochromatic 5100 \AA, and the integrated luminosity for the
   broad component of \hbeta. The spectral parameters are also a
   fundamental key to obtain a reliable classification.
   Spectroscopic observations with a resolution of 4.5 \AA\ were
   performed for a sample of eight galactic nuclei extracted from
   the Second Byurakan Survey, and one companion galaxy of
   SBS~1204+505B. The EW and FWHM of the emission lines were
   measured, and useful line ratios and diagnostic diagrams were
   used for object classification and reddening estimates.
   Intrinsic reddening quantities were calculated for all non QSO,
   i.e. seven objects. Particularly the amount of reddening, \bv\
   color excess, extinction coefficient and optical depths in the
   $V$ band, at \halpha\ and \hbeta\ wavelengths, and at 5100 \AA.
   The broad line region size was also estimated for seven objects,
   as well as the central black hole masses. Three mass estimates
   were usually performed for each object and compared. A peculiar
   line, probably \hei\ $\lambda 5048$, is detected in the QSO
   SBS~1626+554. Evidence for a stratified narrow line region is
   found for the two narrow line Seyfert 1 galaxies included in the
   sample. A revised classification is proposed for two objects, and
   the companion galaxy of SBS~1204+505B is reported as a nuclear
   starburst galaxy.
\end{abstract}




\section{Introduction} \label{sec:introd}


  The Second Byurakan Survey (SBS) is an objective prism survey of
  $991^{\sq}$ carried out in the 1-m Schmidt telescope of the
  Byurakan Observatory \citep{stepa05}. The catalogue is complete at
  85\% for AGNs brighter than $B=17.5$, from which it includes 761
  objects: 155 Seyfert (Sy) galaxies, 596 quasi-stellar-objects
  (QSO) and 10 BL Lac objects.

  A considerable effort has been made to classify all the AGNs
  listed in the SBS. Follow-up spectroscopic observations were
  carried out with the 6~m telescope of the Special Astrophysical
  Observatory (SAO, Russia), the 4.5~m Multi-mirror Telescope (MMT,
  USA), the 2.6~m telescope of the Byurakan Observatory (Armenia),
  and the 2.1~m telescopes of the Guilermo Haro Observatory and the
  National Astronomical Observatory in San Pedro Martir (both in
  Mexico). As a rule, these follow-up observations had spectral
  resolutions of 5-6 or 10-11~\AA\ over the spectral range of
  3300-9000~\AA\ \citetext{see \citealp{stepa05} for further
  details}. From the 155 Sy galaxies, 38 are classified as broad
  line Sy~1, 31 as narrow line Sy~1 (NLS1), 25 as Sy~1.5, 8 as
  Sy~1.9, 44 as Sy~2. The catalogue also includes 90 low ionization
  nuclear emission-line regions (LINERs), 562 starburst nuclei (SBN)
  and starburst galaxies (SB), 195 blue compact dwarf galaxies, and
  150 \hii\ galaxies


  AGNs are classified in two types according to their optical
  spectra. Type~1 comprises the objects with broad permitted
  emission lines, and narrow permitted and forbidden lines. Type~2
  consists of the objects with only narrow lines. However, in the
  paradigm of the Unified Model \citep[see review by][]{antonuc93},
  the power source and physical conditions are thought to be
  inherently similar for both types of objects. The differences may
  arise from the amount of the intrinsic energy output, the angle of
  vision, and the contribution from luminous young stars.


  The intrinsic energy output or luminosity depends on the physical
  conditions in the inner part of the AGN. Two main quantities
  determine the energy production rate: the mass of the central
  black hole (BH) and the accretion rate. Other variables, such as
  the BH angular momentum or the structure of the accretion disk,
  are of secondary importance and may be used to fine-tune the
  theoretical models. In the last years a strong correlation between
  the BH mass and the host-galaxy bulge velocity dispersion has been
  found for both inactive and active galaxies \citep{gebhardt,
  ferra00, tremaine}, which confirms the impact that the presence of
  the central BH has on the galaxy evolution. At the same time,
  reverberation mapping studies have uncovered empirical relations
  between the size of the broad line region (BLR) and different
  estimates of the AGN luminosity \citep{peterson93, kaspi05}.


  According with the Unified Model, the viewing angle determines the
  observed properties of the AGN. Leaving relativistic-jet effects
  in radio loud AGNs aside, the Unified Model anticipates that the
  inner nuclear regions of AGNs are embedded inside dusty tori. The
  detection of broad emission lines in the polarized light of the
  Sy~2 galaxy NGC~1068 by \citet{antonuc85} was a very convincing
  evidence that Sy~1 and Sy~2 galaxies are basically the same
  objects seen from different angles. Depending on the angle of
  vision, the torus may obscure the inner region. Type~1 objects are
  thus unobscured AGN, while Type~2 are obscured.


  Classifying Type~1 AGN in different categories is to some extent
  less complicated than classifying Type~2 objects. The broad bands
  arise exclusively from the BLR of Type~1 AGNs. This has lead to
  more phenomenological studies of the line profiles \citep[see
  review by][]{sulentic}. In contrast, the lines arising from the
  narrow line region (NLR) often compete with the lines emitted from
  \hii\ regions-like, making difficult the interpretation of the
  theoretical results derived from photoionization models. Evidence
  is growing that nuclear and star formation activities may have a
  common trigger, which complicates the classification process. In
  reality, many objects are probably misclassified because of poor
  quality and low resolution spectroscopic data \citep{veron},
  problems to separate the broad and narrow components of the Balmer
  lines, and the presence of \feii\ blends that interferes with the
  measurements of the intensity and the profile parameters of
  \hbeta.


  This paper initially presents spectral observations of eight SBS
  objects and one companion galaxy of SBS~1204+505B to investigate
  the intrinsic reddening, the central BH mass, the size of the BLR,
  and to review their classification. The term \emph{companion} will
  be used hereafter exclusively to designate the companion galaxy of
  SBS~1204+505B reported in this paper. Any other galaxy which may
  be related with other SBS objects will be designed as a
  \emph{neighbor}. The observations were performed as part of a
  follow-up program to obtain medium resolution, high signal to
  noise observations of the AGNs catalogued in the SBS. The
  remaining of the paper is arranged in the following order: \S
  \ref{sec:observ} description of the observations; \S
  \ref{sec:reduct} data reduction; \S \ref{sec:results} presentation
  of the general results; \S \ref{sec:objects} individual analysis
  of each object; and \S \ref{sec:conclusions} a summary of the main
  topics discussed.


\section{Observations} \label{sec:observ}


  Spectroscopic observation with the 2.1-m telescope of the National
  Astronomical Observatory in San Pedro Martir (Mexico) have been
  carried out in April 11--14 of 2002, with the Boller \& Chivens
  spectrophotometer, equipped with a SITE $1K\times1K$ pixel CCD
  installed at the Cassegrain focus. Two spectral ranges were
  covered for most objects: 4800-7200~\AA\ (or 5150-7200~\AA) and
  6500-8000~\AA, the only exception being SBS 0944+540 which was
  observed in the range between 6800-9000~\AA. The grating used was
  a 600~l~mm$^{-1}$ with blaze angle $8 \fdg 63$, which yields a
  dispersion of 94~\AA~mm$^{-1}$ at the front of the CCD. The slit
  width was $2 \farcs 23$, resulting in an effective instrumental
  spectral resolution of 4.5~\AA\ which coincides with the
  instrumental broadening measured from the night sky lines. Two
  1800~s exposures were obtained in each waveband for all objects.
  The continuum of each object was detected with a signal-to-noise
  (S/N) between 30 and 50.


  Table~\ref{tab:log} contains the log of observations. The columns
  list: (1) the SBS designation (equinox B1950), according with the
  IAU nomenclature (except the object referred as
  \textit{companion}, which would be SBS~1204+505A but it is not
  included explicitly in the SBS catalogue); (2)-(3) the J2000.0
  coordinates with an accuracy of $\pm 1\arcsec$; (4) $B$ magnitude
  as listed in the SBS catalog, with an accuracy of about $\pm 0 \fm
  5$ (one decimal place) or $\pm 0 \fm 05$ (two decimal places),
  except for SBS~1204+505B and its companion galaxy, for which the
  magnitudes were calculated from the Sloan Digitalized Sky Survey
  (SDSS, Data Release 5) photometric database (see the procedure
  used in the description of these objects in \S
  \ref{sec:objects}); (5) the redshifts; (6) absolute magnitudes
  $M_{B}$ (see eq. \ref{eq:absb} below); (7) the dates of
  observation; (8) the spectral ranges covered; and (9) the AGN
  spectral types.

  The absolute magnitude $M_{B}$ in Table~\ref{tab:log} was
  calculated from the expression:

   \begin{equation}\label{eq:absb}
    M_B = B - 5\,\log\,\biggl[\,z\left(1+\frac{z}{2}\right)\biggr]
    + 2.5\,(1-\alpha)\; \log\,(1+z) - 43.01,
   \end{equation}

  \noindent for q$_{o}=0$, H$_{0}=75$~\kms~Mpc$^{-1}$. A spectral
  index value of $\alpha$=0.7 (for a power law of the form
  $F_\nu\propto\nu^{-\alpha}$) was adopted for the K correction,
  which is negligible for the low redshift objects ($z<0.07$).

\subsection{Classification criteria}

  The objects have been classified according with the spectroscopic
  criteria outlined in \citet{stepa05}. The following is a short
  description of these criteria which includes the galactic nuclei
  types considered in this paper (namely QSO, Sy galaxies, and SBN)
  along with other related AGNs (for a complete description of AGN
  classification criteria see \citealt{stepa05}):
  \begin{itemize}
    \item QSO - Star-like visual images and very broad permitted
    emission lines. Usually these lines have a Full Width Half
    Maximum (FWHM) larger than 5000 \kms. The absolute magnitude
    $M_B = -23$ separates the QSOs from other AGNs in the SBS
    \citep[see fig. 22 of][]{stepa05}.
    \item Sy~1 - AGNs with broad permitted Balmer lines and narrow
    forbidden lines. The FWHM of the Balmer broad lines is usually
    in the range between 1000 and 6000 \kms. The FWHM of the
    forbidden narrow lines are in the range between 300 and 1000
    \kms\ \citep[e.g.][]{oster76}.
    \item NLS1 - These AGNs where described by \citet{oster85}.
    They have narrow permitted lines only slightly broader than the
    forbidden lines (FWHM of \hbeta\ smaller than 2000 \kms;
    \citealt{good}). The ratio \oiii $\lambda 5007$/\hbeta\ smaller
    than 3, but exceptions are allowed if there are strong emission
    lines of \fevii\ and \fex. Many NLS1 have been discovered in
    X-rays by ROSAT and they have generally steeper soft X-ray
    continuum slopes than normal Sy 1s. Many NLS1 also show rapid
    soft X-ray variability.
    \item Sy~1.5 - AGNs which have a discernible narrow \ion{H}{2}
    profile superposed on a broad component \citep{oster76}.
    \item Sy~1.8 - AGNs which have relatively weak broad \halpha\
    and \hbeta\ components, superimposed on a strong narrow
    component.
    \item Sy~1.9 - AGNs which have a relatively weak broad \halpha\
    component, superimposed on a strong narrow component. The broad
    \hbeta\ component is not seen.
    \item Sy~2 - AGNs which have strong narrow components, but not
    broad components. A secondary criterion is that the \oiii
    $\lambda 5007$/\hbeta\ ratio must be equal or larger than 3.
    \item LINER - Low Ionized Nuclear Emission-line Region. LINERs
    are narrow line, low activity AGNs with line ratios \oii
    $\lambda 3727$/\oiii $\lambda 5007 \geq 1$, and \oi $\lambda
    6300$/\oiii $\lambda 5007 \geq 1/3$ \citep{heckman}.
    \citet{kauffmann} also propose a line ratio \nii $\lambda
    6584$/\halpha $\geq 0.6$. \citet{ho} have detected a weak broad
    component in some of these objects.
    \item SBN - These objects are spiral galaxies with a bright,
    blue nucleus which emits a strong narrow emission line spectrum
    similar to low-ionization \hii\ regions. \citet{balzano}
    proposed three main criteria to separate SBN from other
    starburst objets: (a) strong, narrow (FWHM$\leq 250$ \kms), low
    ionization (with \oiii/\hbeta$<$3) emission lines; (b) absolute
    bolometric magnitudes between -17.5 and 22.5; (c) conspicuous
    stellar nuclei.
  \end{itemize}

\section{Data Reduction} \label{sec:reduct}


  Standard data reduction procedures were applied to the
  observations, i.e. bias subtraction, flat field, illumination and
  response corrections, cosmic ray removal, wavelength
  linearization, atmospheric extinction correction, and flux
  calibration, by using the IRAF package \textit{doslit}.

  The wavelength calibration was accomplished with He and Ar lines
  from a standard comparison lamp observed before and after each
  object, and with the telescope still pointing at the object to
  avoid changes in the calibration due to structural bending. All
  the spectra were flux-calibrated with standard stars observed
  throughout the night. Fig. \ref{fig:spec} shows the reduced
  spectra for the SBS objects.


  \citet{cardelli} and \citet{odonnell} have proposed the use of
  different polynomial empirical fits (up to order eight) for
  different wavelength ranges to model the galactic extinction law.
  As shown by the later author \citep[see figs. 2 and 3
  in][]{odonnell}, the polynomial functions have a tendency to
  introduce spurious bumps and depressions depending on the order of
  the polynomial fit (not to mention the wild behavior of high order
  polynomials outside of the fitting range).

  NED implements a procedure to calculate the absorption in the
  visible and infrared bands based on \citet{schlegel}, who use the
  procedures proposed by \citet{cardelli} and \citet{odonnell}.
  Given the data provided by NED, a simpler fit than the high-order
  polynomials proposed by these authors, or even data interpolation,
  would be enough for correcting for galactic extinction in the
  limited range of the spectroscopic observations such as the ones
  presented in this paper.

  Values for the galactic extinction estimates were obtained from
  NED as total absorption $A_{\lambda}$ in magnitudes for the
  $UBVRIJHKL'$ bands. For convenience, we will perform the galactic
  extinction correction in flux rather than in magnitude units.
  Given the absorption $A_{\lambda}$, it is trivial to calculate the
  transmittance $T_{\lambda}$ which is defined as the ratio of the
  transmitted to the incident radiant power:

  \begin{equation}\label{eq:transmit}
  \centering
  T_{\lambda} = 10^{-0.4A_{\lambda}}.
  \end{equation}

  The resulting transmittance data can be fitted very well by an
  exponential associate function:

  \begin{equation}\label{eq:exasf}
  T_{\lambda} = b_0 + b_1 (1-e^{-\lambda/\lambda_1}) + b_2
  (1-e^{-\lambda/\lambda_2}),
  \end{equation}

  \noindent were $b_0$, $b_1$, $b_2$, $\lambda_1$ and $\lambda_2$
  are the fitted parameters. With increasing wavelength, the
  transmittance reaches a constant (asymptotic) value of
  $T_{\infty}=1$, hence the linear parameters have the constraint
  that:

  \begin{equation}
  b_0+b_1+b_2=1.
  \end{equation}

  The exponential associate function produces reasonable good fits
  for $T_{\lambda} > 0.1$; this is a much lower value than the
  transmittances calculated for the objects included in the SBS
  catalog, which are located well above the galactic plane and have
  small amounts of galactic extinction ($T_{U}>0.7$). As an example,
  Fig.~\ref{fig:tranmit} shows the exponential associate fit of the
  transmittance data as calculated for SBS~1136+594.


  Different authors have shown their preferences for either Gaussian
  or Lorentzian profiles to fit the broad emission lines. The choice
  usually depended on the properties or the kind of AGN being
  considered. As a few examples, it is worth to cite
  \citet{marziani}, for whom Lorentzian functions provide a better
  fit for sources with a high luminosity to mass ratio, while low
  luminosity to mass ratio sources are best fitted by a double
  Gaussian; \citet{dumont90b}, \citet{moran}, \citet{leighly}, and
  \citet{veron} find that many NLS1 galaxies have symmetric emission
  lines with Lorentzian rather than Gaussian profiles. On the other
  hand, \citet{rodriguez} argue that Lorentzian profiles are not
  suited to represent NLS1 broad emission lines. The broad emission
  lines analyzed in this paper have been fitted using both Gaussian
  and Lorentzian profiles.

  Some objects show evidence of very broad extended wings to the
  Balmer lines. These wings do not disappear using different
  continuum levels and can be approximated by a logarithmic law
  \citep{blumenthal}.

  The narrow emission lines are well fitted using single Gaussian
  profiles. In some cases, the wings show a minor excess near the
  base and a Lorentzian profile can be fitted yielding a FWHM
  estimate slightly lower than the Gaussian fit, in accordance with
  \citet{veron}. However, the values reported in this paper are
  always calculated using Gaussian profiles.

  Strong \halpha\ and \nii\ blends were fitted simultaneously
  allowing the line ratio \linebreak \nii$\lambda 6584$/\nii$\lambda
  6548$ to vary (within the expected errors) between 2.7 and 3.3,
  instead of forcing to fulfill the theoretical value of 2.94. The
  same procedure was used to fit the \oiii\ $\lambda \lambda
  4959,\,5007$ lines which have a theoretical value of 2.98
  \citep{storey}. The measured line ratios were compared with the
  theoretical values to check for possible unmodelled errors in the
  fitted spectra (particularly for objects modelled using several
  components for the Balmer lines, or with a strong emission of
  \feii).


  In order to detect \heii\ $\lambda 4686$ and to fit the real
  \hbeta\ profile, \feii\ subtraction was performed using a template
  obtained from the NLS1 galaxy IZw1 \citetext{P. Marziani, priv.\
  comm.} whose spectrum is widely used as an \feii\ template. The
  procedure to subtract the \feii\ is the same as the one reported
  by \citet{zamanov}.


\section{Results} \label{sec:results}


  Table~\ref{tab:widths} shows the rest frame equivalent width (EW)
  in $\AA$ and the full width half maximum (FWHM) in \kms\ for the
  observed lines in each object. These quantities have been
  calculated using the following expressions:

  \begin{equation}
    \textrm{EW} = \frac{F_l}{(1+z)F_c},
  \end{equation}

  \noindent where $F_l$ and $F_c$ are the observed fluxes of the
  line and the continuum, respectively, and

  \begin{equation}
    \textrm{FWHM} = \frac{\textrm{c}}{1+z} \frac{\Delta\lambda_{obs}}{\lambda_e},
  \end{equation}

  \noindent where c is the speed of light, $\Delta\lambda_{obs}$ is
  the observed FWHM in $\AA$, and $\lambda_{e}$ the rest frame
  wavelength of the line.

  The first column in Table~\ref{tab:widths} identifies the emission
  line, and columns 2-19 alternatively the EW in \AA\ and the FWHM
  in \kms\ for each object. In some cases, the spectra show a small
  feature that strongly suggests the existence of a known line, but
  the S/N ratio is below the detection level ($3 \sigma$). For these
  features, the parameters of the fit are not accurate, and
  therefore their EW are indicated as superior limits, and no FWHM
  value is available.

  The fluxes for the iron blends between 4400-4680 \AA\ (\feii\
  $\lambda 4540$), and 5100-5500 \AA\ (\feii\ $\lambda 5300$) have
  been calculated from the iron template. The FWHM data were
  corrected for instrumental broadening (4.5 \AA).


  The discussion will be centered on the Gaussian fits, although the
  parameters of the Lorentzian fits were also calculated for all the
  permitted lines. Table~\ref{tab:lore} lists the Lorentzian
  parameters for those object for which the Lorentzian fits
  reproduce the Balmer profiles using less components than the
  Gaussian fits. In these cases, a single component was used except
  for the QSO SBS~0944+540. For this object, the simplest accurate
  fit was obtained using two broad Lorentzian components. Otherwise,
  the structure of Table~\ref{tab:lore} is similar to
  Table~\ref{tab:widths}.

\subsection{Narrow lines and diagnostic diagrams}


  From Table~\ref{tab:widths} it is evident that for most objects
  there are manifest differences between the FWHM of the lines found
  in the \hbeta\ and \halpha\ regions (SBS~0848+526, 1118+541,
  1204+505B and its companion, and 1340+569), and between the narrow
  components of the permitted lines and the forbidden lines FWHM
  (SBS~1118+541 and 1136+595), and even among the forbidden lines
  (SBS~1118+541). The correlation between the line widths and both
  the critical density and the ionization potential are well known
  effects that can explain these differences \citep[e.g.][]{pelat,
  filip84, derobert, espey}. \citet{filip85} find that \oi $\lambda
  6300$ in many LINER galaxies was broader than the \sii $\lambda
  \lambda 6716,\,6731$ lines, which they explain in terms of the
  different critical densities $n_e$(crit) for collisional
  de-excitation for these lines. Following this line of thought, the
  critical densities for the \fevii\ and \fex\ ions are $3.6 \times
  10^7 \rm{cm}^{-3}$ and $4.8 \times 10^9 \rm{cm}^{-3}$,
  respectively. These critical densities are much higher than the
  mean density in the NLR, and the former lies even close to the
  density in the BLR. The high critical densities may explain the
  large FWHM values measured for these lines in some objects
  (SBS~1118+541 and 1340+569; see Table~\ref{tab:widths} and the
  discussion in \S \ref{sec:objects}).

  The widths of the narrow lines are probably virial in their
  origin. For example, \citet{nelson96} have found a very good
  correlation between the width of the \oiii\ and the stellar
  velocity dispersion within the bulge. The virial origin for the
  width of the narrow lines, along with the correlation between
  these widths and both the critical density and the ionization
  potential reviewed above, imply a radial stratification of the
  NLR.


  Table~\ref{tab:ratios} lists the emission line ratios for objects
  that show clear narrow Balmer components. For the objects
  classified as QSO, no narrow component was necessary to fit the
  Balmer lines, thus they are not included in the table. The line
  ratios have been corrected from galactic extinction. Column~1
  identifies the SBS object, and columns~2-6 list the corresponding
  line ratio with respect to the narrow component either of \halpha\
  or \hbeta. Detection limits for some line ratios involving
  forbidden lines (\oiii, \oi\ and \nii) have been calculated and
  plotted in the diagnostic diagrams (see below). Note that for
  SBS~0848+526 and SBS~1340+569, the \sii\ $\lambda 6717$ line is
  outside the spectral range of the observations.


  Emission line models for AGNs have lead to search for diagnostic
  line-intensity ratios to analyze the observed optical spectra of
  AGNs and separate them from star-forming regions.
  \citet{baldwin81} proposed a set of empirical diagnostic diagrams
  which, after revisions by \citet{oster85} and \citet{veilleux},
  are the most currently used. The physics behind these diagrams is
  that the presence of free electrons generated by X-ray
  photoionization in partially ionized zones will enhance the
  strengths of the forbidden low-excitation lines (particularly \oi,
  \sii\ and \nii) produced by collisional excitation with respect to
  Balmer lines in narrow line AGNs, in comparison with \hii\ region
  nuclei. Besides, \ion{O}{3} is produced by hard ultraviolet
  photons, relatively more abundant in the AGN power law spectra
  than in \hii\ regions. Thus, \oiii\ is an indicator of the mean
  level of ionization and temperature, and it is enhanced by
  photoionization from the harder AGN continua. This makes of
  \oiii/\hbeta\ an excellent indicator of the AGN activity, and all
  the diagnostic diagrams in the optical range include this line
  ratio.

  \citet{kewley01,kewley06} have derived the first purely
  theoretical classification scheme. The relevant narrow line ratios
  for diagnostic diagrams are plotted in Fig.~\ref{fig:diagnos},
  along with graphs generated from \citeauthor{kewley01}

\subsection{Reddening}

  Solving the case B equation for the equilibrium level population
  at temperature $10^4$~K and electronic density $10^4\
  \rm{cm}^{-3}$ for \hii\ regions yields a ratio for the Balmer
  recombination decrement \halpha/\hbeta\ of 2.85 \citep{oster89}.
  \citet{gaskell84} has shown that this ratio depends weakly on the
  electron density, but on the metallicity and the presence of a
  strong X-ray continuum. The intrinsic value for the NLR has been
  proposed to be 3.1 \citep{gaskell82,gaskell84,halpern}, and larger
  observed values are imputed to reddening by dust. As it is usual
  in AGN, all the \halpha/\hbeta\ line ratios shown in
  Table~\ref{tab:ratios} are larger than the fiducial and proposed
  values of 2.85 and 3.1, respectively. However it is worth to
  remark that dust might be not the only responsible for large
  \halpha/\hbeta\ line ratios. \citet{netzer} has argued that
  transfer effects, self-absorption and collisional excitation of
  hydrogen lines are important in the BLR (which renders the broad
  lines unreliable for reddening measurements) and they may also
  have observable consequences in the NLR, affecting the
  $\rm{Ly}\alpha$/\hbeta\ and \halpha/\hbeta\ ratios
  \citep{binette}.

  \citet{witt} have shown that heavily reddened components may have
  a negligible contribution to the total reddening, and that
  'bluing' due to scattering partially compensates for reddening by
  extinction. Thus, to estimate the intrinsic reddening by dust
  within the host galaxy, we will assume that the local reddening
  has a continuous distribution along the line of sight. Besides, we
  will also assume that the optical properties of the dust in
  emission line galaxies are the same than the properties of dust in
  the solar neighborhood \citep{savage}. Thus, the magnitudes for
  the unreddened and the observed fluxes ($I$ and $F$, respectively)
  are related by:

  \begin{equation}\label{eq:magred}
    -2.5 \log I_{\lambda} = -2.5 \log F_{\lambda} - C\ \frac{A_{\lambda}}{A_{V}}
  \end{equation}

  \noindent where $C$ depends on the reddening and $A_{\lambda}/A_V$
  is the extinction coefficient with $V$ normalization.

  Applying eq.~\ref{eq:magred} to the observed and intrinsic fluxes
  for \halpha\ and \hbeta, we obtain:

  \begin{equation}\label{eq:constant}
    C = 2.5\ \Big( \frac{A_{\beta}-A_{\alpha}}{A_V} \Big) ^{-1}\
        \Big( \log \frac{F_{\alpha}}{F_{\beta}} - \log \frac{I_{\alpha}}{I_{\beta}}  \Big)
  \end{equation}

  \noindent where $A_{\alpha}/A_V = 0.80$ and $A_{\beta}/A_V = 1.19$
  (we will also adopt $A_B/A_V = 1.33$ and $A_{5100 \AA}/A_V = 1.11$
  hereafter). To calculate $C$, we assume the intrinsic value of
  $I_{\alpha}/I_{\beta} = 3.1$ for the unreddened \halpha/\hbeta\
  ratio.

  Other interesting parameters can be also derived, such as the \bv\
  color excess $E_{B-V}$ (recall that $A_B/A_V$ is the given
  quantity, rather than $A_B$ and $A_V$), the absorption coefficient
  $A_V$ \citep{whittet}, and the optical depth $\tau_{\lambda}$ at
  any wavelength:

  \begin{eqnarray}
    E_{B-V} &=& C\ \Big( \frac{A_B}{A_V}-1 \Big)  \label{eq:colorex}, \\
    A_V &=& 3.05\ E_{B-V} \label{eq:extin}, \\
    \tau_{\lambda} &=& \frac{A_{\lambda}}{2.5 \log e}  \label{eq:opdepth}.
  \end{eqnarray}


  Table~\ref{tab:redden} lists all the significant reddening
  parameters of the NLR for the objects presented in this paper. The
  columns indicate: (1) the object designation; (2) the reddening
  constant $C$ (eq. \ref{eq:constant}); (3) the color excess
  $E_{B-V}$ (eq. \ref{eq:colorex}); (4) the extinction in the $V$
  band (eq. \ref{eq:extin}); and (5), (6) (7) and (8) the optical
  depths (eq. \ref{eq:opdepth}) for the $V$ band, \halpha, \hbeta\
  and at $\lambda = 5100\ \AA$.

\subsection{Mass of the central black hole}

  Kinematic and reverberation mapping studies have lead to the
  finding of a strong correlation between the central black hole
  masses ($M_{BH}$) and the bulge stellar velocity dispersion
  ($\sigma$). This correlation has been studied by several authors
  and it has been shown to be valid for both, AGN and non AGN
  galaxies \citep[e.g.][]{ferra00, ferra01, gebhardt}.
  \citet{tremaine} have proposed the expression:

  \begin{equation}\label{eq:tremaine}
    M_{\rm{BH}} = 10^{8.13}\ \Bigl( \frac{\sigma}{200\ \rm{km\ s}^{-1} } \Bigr)^{4.02} \textrm{M}_{\odot}.
  \end{equation}

  \citet{nelson96} and \citet{nelson00} assumed that for most AGN
  the forbidden line kinematics is dominated by virial motion in
  the host galaxy bulge, finding that the bulge velocity dispersion
  can be estimated from the width of the \oiii\ line:

  \begin{equation}\label{eq:voiii}
    \sigma = \frac{\textrm{FWHM}(\textrm{[\ion{O}{3}]})}{2.35}.
  \end{equation}

  However, there is evidence that the use of FWHM(\oiii) as an
  estimator of the bulge stellar velocity dispersion in AGN tends to
  produce large scatters and may fail for individual objects
  \citep[see discusion by][]{onken}.

  On the other hand, \citet{kaspi00} monitored a sample of 28
  Palomar-Green quasars for reverberation mapping and found a
  correlation between the size of the BLR and the luminosity at
  $\lambda = 5100$~\AA, which was later updated \citep{kaspi05} to
  the relation:

  \begin{equation}\label{eq:rblr1}
      R_{\rm{BLR}} = 22.3\ \Bigl( \frac{\lambda L_{\lambda}}{10^{44}\ \textrm{erg s}^{-1} }  \Bigr) ^{0.69} \textrm{light days}.
  \end{equation}

  \noindent This expression relies on the AGN continuum
  monochromatic luminosity at 5100~\AA\ which is a good
  approximation in the case of QSO, although for other AGN the host
  galaxy contribution at this wavelength can be very important.
  \citet{kaspi05} propose other possible BLR size and luminosity
  relations. In particular, the relation with the luminosity of the
  \hbeta\ broad component seems to be as accurate as with the
  luminosity at 5100~\AA\ and both power law fits share the same
  exponent. Furthermore, it has the advantage that the estimate for
  the luminosity of the broad component of \hbeta\ is not
  contaminated by the host galaxy. Following the same procedure as
  \citet{kaspi05} to obtain the parameters in eq.~\ref{eq:rblr1}
  (i.e. averaging the parameter values obtained from the linear
  regression method and from the bivariate correlated errors and
  intrinsic scatter), we obtain an expression that relates the BLR
  size with the luminosity of the broad component of \hbeta:

  \begin{equation}\label{eq:rblr2}
    R_{\rm{BLR}} = 85.6\ \Bigl( \frac{L_{\textrm{\hbeta}}}{10^{43}\ \textrm{erg s}^{-1} }  \Bigr) ^{0.69} \textrm{light days}.
  \end{equation}

  \noindent Assuming that the \hbeta\ widths indicate the random
  orbits of the BLR material moving with Keplerian velocity:

  \begin{equation}\label{eq:vhb}
    V = \frac{\sqrt{3}}{2}\ \textrm{FWHM}(\textrm{\hbeta})
  \end{equation}

  \noindent the black hole mass can be expressed as:

  \begin{equation}\label{eq:mhb}
    M_{\rm{BH}} = R_{\rm{BLR}}\ V^{2}\ G^{-1}
  \end{equation}

  \noindent where $G^{-1}$ is the gravitational constant.

  Note that \citet{onken} has proposed that the kinematics and
  geometry of the BLR introduce a scaling factor of $f=5.5$ multiplying
  the mass. This factor has not been introduced in the analysis.

  Table~\ref{tab:masses} lists the relevant parameters for the black
  holes. Column (1) identifies the object; (2) and (3) show the
  logarithm of the luminosities in solar units of the continuum at
  5100 \AA\ and for the \hbeta\ broadest line component,
  respectively, corrected by galactic extinction and intrinsic
  reddening when available; (4) and (5) the size of the BLR in light
  days calculated from the luminosities at 5100~\AA\ and from the
  broadest \hbeta\ component (eqs. \ref{eq:rblr1} and
  \ref{eq:rblr2}), respectively; (6), (7) and (8) the logarithms of
  the mass of the black hole in solar units; the first two estimates
  were calculated through the size of the BLR listed in columns (4)
  and (5) (eq. \ref{eq:mhb}), and the orbital velocity of the BLR
  clouds estimated from the width of the broadest \hbeta\ component
  (eq. \ref{eq:vhb}); the last mass was estimated from the width of
  the \oiii\ (eqs. \ref{eq:tremaine} and \ref{eq:voiii}); finally,
  column (9) compares the mass estimates listed in columns 6 and 8.

  From the values listed in Table~\ref{tab:masses}, the BH masses
  calculated from the kinematics (FWHM) of the broadest component of
  \hbeta\ and the energy output of the BLR, using either the
  monochromatic luminosity at 5100~\AA\ or the luminosity of the
  broadest \hbeta\ component \citep{kaspi05}, are very consistent
  (the largest discrepancy, for SBS~1340+569, is only a factor 2.5).
  However, the comparison of these two estimates with the BH mass
  calculated from the velocity dispersion in the galactic bulge
  \citep{tremaine}, estimated from the FWHM(\oiii), shows
  significantly larger discrepancies (a factor of 44 for
  SBS~1136+594), as displayed in the last column of
  Table~\ref{tab:masses}.


\section{Notes on individual objects} \label{sec:objects}

\subsection{\protect\objectname[]{SBS 0848+526}}

  This emission line object is an X-ray (ROSAT), radio (FIRST) and
  infrared (2MASS) source. The host galaxy is aligned roughly in the
  NS direction, and it has a brighter neighbor (around 0~\fm 5)
  separated by 14\arcsec\ to the NE (see description below). The
  SDSS image shows that the two objects have bright nuclei, and a
  plume extending between both galaxies, approximately aligned in
  the EW direction, which coincides with the slit alignment during
  the observations. Thus, it was possible to detect extended
  emission of \halpha, \hbeta\ and \oiii\ along this plume. Hence,
  the pair of objects seems to be another case of nuclear activity
  (either star formation and/or AGN) triggered by galaxy
  interaction.

  The spectra obtained of the SBS object show only narrow lines. It
  is worth noticing that the \halpha/\hbeta\ ratio for this object
  is large (see Table~\ref{tab:ratios}), which indicates a
  substantial reddening, with an intrinsic extinction coefficient of
  $A_V=2.22$ (see Table~\ref{tab:redden}).

  Besides \halpha\ and \hbeta, the permitted \hei\ $\lambda 5816$
  line is also observed. For the forbidden lines, the spectrum shows
  the \oiii\ $\lambda \lambda 4959,\,5007$ and the \nii\ $\lambda
  \lambda 6548,\,6584$ lines, as well as a weak \oi\ $\lambda
  6300$. The luminosity of this SBS object is M$_B$=-19.9, the line
  ratio \oiii/\hbeta=1.41, and the FWHM$< 250$~\kms\ for all the
  emission lines. Hence, following \citet{balzano}, the object is
  correctly classified in the SBS catalogue as a SBN by its optical
  properties. In Fig.~\ref{fig:diagnos} we see that the available
  line ratios for this object also agree with an \hii\ region-like.

  The classification of SBS~0848+526 as a SBN is also compatible
  with the soft X-ray emission detected by ROSAT. However, AGNs also
  emit in soft X-rays, which sometimes can be detected even in the
  case of significantly reddened objects. Recently
  \citet{jim03,jim05} have studied the contribution of nuclear
  starbursts to the X-ray emission from AGNs. Different approaches
  to disentangle the origin of the X-ray emission consist on careful
  modelling of the starburst and AGN emissions, higher X-ray imaging
  resolution than the data available from ROSAT All Sky Survey
  (25\arcsec), and hard X-ray observations.

  As mentioned above, the SBS object has a neighbor galaxy. This
  galaxy is also an X-ray source (ROSAT). It is not included in the
  sample presented in this paper, but a few properties may be easily
  inferred from the SDSS data. The SDSS spectrum of the neighbor
  galaxy shows several emission lines (Balmer series, \feii\ blends,
  \oiii, \oi, \nii-\halpha\ blend, and \sii) at the same redshift
  (within the errors) as SBS~0848+526. This spectrum also shows the
  presence of the \feii\ blends, which are usually associated with
  the accretion disk \citep[e.g.][]{dumont90a, dumont90b}. Besides,
  \halpha\ and \hbeta\ show evidence for a very weak broad
  component. Thus the most probable classification for this
  neighbor is as a Sy~1.8 galaxy.

\subsection{\protect\objectname[]{SBS 0944+540}}

  This is a well known QSO \citep[e.g.][]{veron,hewitt} and a radio
  weak source \citep{bischof}. It shows strong emission of \feii\
  blends. After subtracting the \feii\ emission, there is evidence
  for a weak \heii\ $\lambda 4686$ emission. Although the spectrum
  at the red end is noisy, it shows a weak increase near $8750~\AA$,
  which corresponds to the \hei\ $\lambda 5876$ emission line. This
  line is clearly detected on the SDSS spectrum. Neither forbidden
  lines nor narrow components of permitted lines are observed.

  The \hbeta\ profile is very asymmetric and shows a strong blue
  bump. This line can be fitted using either two broad Lorentzian or
  three broad Gaussian profiles. In the later case, two broad
  components (FWHM $\approx 2500$~\kms), and a very broad component
  for the extended wings ($\approx 13,000$~\kms) are necessary.
  These wings can also be approximated using a logarithmic law. The
  BH mass estimated from the widest component of \hbeta\ is
  approximately $6\times 10^9 \rm{M}_\odot$ and the BLR has a
  diameter of 200 light days (see Table~\ref{tab:masses}).

\subsection{\protect\objectname[]{SBS 1118+541}}

  This object is a ROSAT and FIRST source. It has been classified as
  a Sy~1 by \citet{stepa02}, but appears as a NLS1 in the SBS
  catalogue. The permitted lines can be fitted using either Gaussian
  or Lorentzian profiles. In the cases of \halpha\ and \hbeta, two
  Gaussian components are necessary to fit the lines, but a single
  Lorentzian can also reproduce their profiles. The Lorentzian
  profiles for the Balmer lines have a FWHM of 1400-1500~\kms, which
  are in accordance with the NLS1 classification. In the case of the
  Gaussian profiles, the broad components have FWHM in the range
  3000-4000~\kms, while the narrower components are broader than the
  forbidden lines [FWHM(\hbeta n) $\approx 1000$~\kms\ but
  FWHM(\oiii\ $\lambda \lambda 4959,\,5007$) $\approx 540$~\kms].
  \citet{dietrich} studied 12 NLS1 and also found that the Gaussian
  decomposition of \hbeta\ yielded broad components ($\rm{FWHM} =
  3275 \pm 800$~\kms) and intermediate broad component ($\rm{FWHM} =
  1200 \pm 300$~\kms).

  The \heii\ $\lambda 4686$ emission line can be easily
  distinguished and measured after subtracting the intense \feii\
  background. Fig.~\ref{fig:hb1118} shows the spectral region around
  \hbeta\ before and after subtracting the \feii\ blends. The \heii\
  $\lambda 4686$ has acquired special relevance in reverberation
  mapping studies because it responds with negligible delay to
  continuum variations \citep[e.g.][]{ulrich}. With respect to the
  \hei\ $\lambda 5876$ line, it is possible to discern between the
  broad and narrow components. The parameters of the broad component
  (EW and FWHM), which depend on the baseline adopted for the
  subtraction of the continuum, are relatively inaccurate, but the
  narrow component is relatively free of this dependency, and it
  shows a FWHM comparable with \halpha\ and \hbeta.

  The narrow components of the permitted lines are much wider than
  the forbidden lines observed, with the exception of the weak
  \fevii\ $\lambda 6087$ and \fex\ $\lambda 6374$. Thus, the \oiii\
  $\lambda \lambda 4959,\,5007$ and the weak lines of \oi\ $\lambda
  6300$ and \sii\ $\lambda \lambda 6717,\,6731$ have FWHM
  approximately half of the narrow \hbeta\ and \halpha,
  respectively. But the forbidden lines also show a large dispersion
  in their FWHM, from approximately 1000~\kms of \fevii\ to 400~\kms
  of \sii\ $\lambda 6717$. The \nii\ $\lambda \lambda 6548,\,6584$
  lines are not observed, probably blended with the relatively
  intense \halpha\ emission.

  The Lorentzian FWHM, along with the strong \feii\ blends, the
  presence of \fevii\ and \fex, the line ratio \oiii/\hbeta\ $< 3$,
  and the X-ray emission are in accordance with the SBS
  classification as a NLS1. However, it is worth to note that the
  Gaussian analysis shows a \hbeta\ broad component
  (FWHM(\hbeta)=2630~\kms) significantly wider than the nominal NLS1
  limit (FWHM(\hbeta)=2000~\kms). Separating the Balmer lines in a
  broad and a narrow component has lead to a revision of the quality
  of several objects which were previously misclassified as NLS1
  \citep[e.g.][]{veron, botte, bian06a, bian06b}. These
  misclassifications were in part favored by the original
  definition of NLS1, which did not take into consideration the
  impact of a varying mix of the broad and narrow line components
  \citep{oster85, good}.

  The EW of the broad and the narrow components of the Balmer lines
  (Table~\ref{tab:widths}) can be used to compare the proportion of
  the flux arising from the BLR with the flux from the NLR. In this
  case, the BLR output is about twice the emission from the NLR. The
  narrow components \halpha/\hbeta\ ratio (3.71) is moderate for a
  broad line AGN. Assuming a continuous distribution of local
  reddening, this ratio yields an intrinsic extinction coefficient
  of $A_V = 0.42$ and small optical depths which indicate that the
  object is optically thin for dust scattering in the optical bands
  (see Table~\ref{tab:redden}).

  The line ratios lie in the \hii\ region of the diagnostic diagrams
  (Fig.~\ref{fig:diagnos}). \citet{veilleux} have already noted that
  when a Gaussian profile is used for fitting the broad Balmer
  components of NLS1 galaxies, the line ratios show a significant
  spread in the diagnostic diagrams. Hence, rather than an
  indication of SBN activity, this effect is the result of comparing
  the forbidden lines arising from the NLR with the narrow Balmer
  components which in this case originate in an intermediate region
  between the BLR and the NLR.

  The differences in line widths among the forbidden lines as
  compared with the permitted narrow lines indicate a strong
  stratification of the NLR. \citet{wilson} have discussed that the
  FWHM of an emission line approximately represents the bulk motion
  of clouds whose density is equal to the critical density for that
  line \citep[cf.][]{ferguson}. \citet{filip88} used this approximation to show that the NLR
  of the Sy1 nucleus M81 might be stratified, with a dependence such
  that the velocities $v$ of the clouds scale as $v \propto
  n_e^{0.10 \pm 0.03}$. The dependency between the critical
  densities (for a temperature of 10$^4$\,K) and the FWHM of the
  forbidden lines is shown in Fig.~\ref{fig:critdens}. In the case
  of SBS~1118+541 the weighted, linear least-squares fit shows that
  $v \propto n_e^{0.07 \pm 0.01}$.

  Maybe the AGN spatial orientation (NLS1 are presumed to be seen
  face-on) permits to observe the different strata. Besides, the
  small amount of reddening inferred for this object
  (Table~\ref{tab:redden}) does not impede the observation of the
  deeper regions. Of course, such small reddening is also in
  accordance with the face-on view and the presumed distribution of
  the dust in a torus. \citet{ferguson} have shown that that the
  integrated narrow-line spectrum in Sy galaxies can be explained by
  considering an ensemble of clouds in the context of the Locally
  Optimally emitting Cloud (LOC) model \citep{baldwin95}. Under the
  assumption of a Keplerian gravitational velocity field, this model
  can predict the observed line width as a function of the critical
  density.

  The relevant BH parameters for this object are listed in
  Table~\ref{tab:masses}. The mass estimated from the \oiii\ is
  slightly larger but compatible with the mass estimated from the
  broad component of \hbeta, and both are larger than the value
  expected for a NLS1 ($M_{\rm{BH}} \sim 10^7\ \rm{M}_{\odot}$).
  This result counterpoints with \citet{bian04} who have
  investigated the BH-bulge relation in AGN. These authors found
  that, in the case of NLS1, the BH mass estimated from \oiii\ can
  be one order of magnitude larger than the mass estimated from
  \hbeta\ (see their Fig.~2). They used the 150 NLS1 sample
  extracted from the SDSS by \citet{williams}, who developed a
  procedure to measure the width of the whole line halfway between
  the fitted continuum and the line peak from the SDSS spectra.
  Hence, the disagreement is a result of the methodology used to
  measure the FWHM, and can be easily resolved applying Lorentzian
  rather than Gaussian fits. Thus, using the \hbeta\ FWHM listed in
  Table~\ref{tab:lore}, the logarithm of the BH mass in solar units
  calculated from the continuum at 5100~\AA\ changes from 8.19 to
  7.32. This mass is in accordance with the expected value for a
  NLS1 and just an order of magnitude smaller than the mass
  calculated from the \oiii\ FWHM, in agreement with \citet{bian04}.
  The large discrepancy (a factor 7) between the BH masses inferred
  from line fits using Gaussian versus Lorentzian models is
  sobering, showing that much of the energy is in the line wings and
  suggesting that many BH mass estimates presented in the literature
  may have large and probably systematic errors.

\subsection{\protect\objectname[]{SBS 1136+594}}

  This AGN is a ROSAT and 2MASS source. It is classified as a Sy~1.5
  galaxy in the SBS catalogue and by \citet{goncalves}. The \nii\
  lines are very weak (\nii$/$\halpha = 0.2), a characteristic
  already reported by \citet{martel} and \citet{goncalves}
  (\nii$/$\halpha = 0.25 and 0.1, respectively). The \hbeta\ and
  \halpha\ lines show moderate asymmetric profiles
  \citetext{\citeauthor{goncalves} use three components to fit
  \halpha, but only two for \hbeta}.

  Other spectral characteristics are the strong \oiii\ lines which
  are narrower than \hbeta, weak \feii\ emission, the presence of
  \hei\ $\lambda 5876$, \oi\ $\lambda 6300$, \fex\ $\lambda 6374$,
  and \sii\ $\lambda 6717$ at the red end of the observed spectrum.

  Comparing the broad and narrow components of the Balmer lines, the
  flux arising from the BLR is approximately 15 times the flux from
  the NLR, and the line ratios lie in the AGN region of the
  diagnostic diagrams (Fig.~\ref{fig:diagnos}). The large
  \oiii/\hbeta\ line ratio of 10.88 indicates the strength of the
  ultraviolet AGN continuum. However, the \halpha/\hbeta\ ratio
  (7.83) is also the highest in the sample, denoting a large amount
  of intrinsic reddening (see Tables~\ref{tab:ratios} and
  \ref{tab:redden}).

  The \oiii\ FWHM has a value of only 200~\kms. Hence the BH mass
  estimated from the \oiii\ FWHM ($4 \times 10^6~\rm{M}_\odot$) is
  rather humble for a Seyfert galaxy, and it is also a factor
  approximately 44 smaller than the mass estimated from the \hbeta\
  width (Table~\ref{tab:masses}). The size for the BLR, estimated
  from the monochromatic 5100 \AA\ and the \hbeta\ broad component
  luminosities, is large ($\approx 160$ light days) and it is
  comparable to the sizes estimated for QSOs.

\subsection{\protect\objectname[]{SBS 1136+595}}

  This object is a 2MASS source. It was classified as a Sy~1 galaxy
  by \cite{markarian}, but appears as a NLS1 galaxy in the SBS. The
  \feii\ is strong and, after subtraction, it is possible to detect
  a very noisy \heii\ $\lambda 4686$ (Fig.~\ref{fig:hbfit}). The
  FWHM of this line is poorly estimated, yielding a value of
  10240~\kms\ (8700~\kms\ for the Lorentzian fit). This width may
  seem too large when compared with 1490~\kms\ measured for the FWHM
  of \hbeta, but it is worth to remember (see discussion on
  SBS~1118+541 above) that the \heii\ $\lambda 4686$ line shows
  negligible delays with respect to the continuum, and hence
  probably arises from fast moving BLR clouds very close to the
  central BH.

  \halpha\ and \hbeta\ can be fitted using either two Gaussian
  components (narrow and broad) or a single Lorentzian. but for the
  weaker \hgamma\ a single Gaussian can also reproduce the line
  profile. When the Lorentzian profile is applied, the three Balmer
  lines observed have FWHM smaller than 2000~\kms. However, as in
  the case of SBS~1118+541 discussed previously, the broad
  components for the Gaussian fits are much larger than the limit
  for NLS1 galaxies. The remarks about the classification of
  SBS~1118+541 given above are also applicable to SBS~1136+595.

  Also, as in the case of SBS~1118+541, the NLR shows evidence of
  stratification. Thus, the \halpha\ and \hbeta\ narrow components
  have widths much larger than the forbidden lines, and the
  differences among the FWHM of the forbidden lines, in the range
  350-540 \kms, are also large but not as much as in SBS~1118+541,
  probably due to the absence of the \fevii\ and \fex\ lines.
  The narrow components \halpha/\hbeta\ ratio (4.24) and the
  reddening ($A_V=0.88$) are larger than in SBS~1118+541, but still
  moderate for an AGN (Table~\ref{tab:redden}). A comparison between
  the broad and narrow components of the Balmer lines shows that the
  BLR emits $\sim 1.5$ times the NLR. Other lines present in the
  spectrum are \heii\ $\lambda 5876$, \nii\ $\lambda \lambda
  6548-6584$, and \sii\ $\lambda \lambda 6717,\,6731$.

  As in SBS~1118+541, the line ratios also lie in the \hii\ region
  of the diagnostic diagrams (Fig.~\ref{fig:diagnos}), in accordance
  with the result reported by \citet{veilleux} for NLS1 whose broad
  Balmer lines have been fitted with Gaussian profiles.

  The BH mass estimated from the FWHM of the broad component of
  \hbeta\ and \oiii\ agree (Table~\ref{tab:masses}). But again, as
  in SBS~1118+541, the logarithm of the BH mass in solar units
  calculated from the Lorentzian \hbeta\ profile
  (Table~\ref{tab:lore}) and  the continuum at 5100~\AA\ changes
  from 8.06 to 7.17, in accordance with the NLS1 masses reported by
  \citet{bian04}.

\subsection{\protect\objectname[]{SBS 1204+505B} and companion}

  SBS 1204+505B is a ROSAT and 2MASS source. In the SBS catalogue,
  it is classified as a Sy~1.8 galaxy. The object probably belongs
  to a compact group of three galaxies. Relative to the center of
  SBS~1204+505B, one galaxy is roughly to the south and the other to
  the west. The SDSS image from the former galaxy is unlikely to
  correspond to an AGN (no spectrum is available for this galaxy).
  On the contrary, the later galaxy shows a bright stellar-like
  nucleus which indicates some kind of nuclear activity. We refer to
  this galaxy as the \emph{companion}. The spectra for both
  SBS~1204+505B and its companion are also available through the
  SDSS.

  The reported $B$ magnitude of SBS~1204+505B in the SBS catalogue
  is 17.0$\pm 0 \fm 5$. In Table~\ref{tab:log} the magnitude has
  been transformed from the SDSS photometric data ($g'$ and $r'$
  bands) to allow comparison with the companion galaxy. The equation
  for the transformation is borrowed from \citet{smith}:

    \begin{equation}
      B = g' + 0.47(g'-r') + 0.17
    \end{equation}

  \noindent being the SDSS magnitudes $g'$ an $r'$ 16.18 and 15.57
  for the SBS object, and 16.67 and 16.15 for the companion.

  The spectrum of SBS~1204+505B does not show any \feii\ blends. In
  the SDSS spectrum it is possible to distinguish a weak broad
  component of \hbeta, but the spectrum presented in this paper
  shows only a narrow component. \oiii\ $\lambda \lambda
  4959,\,5007$ lines are clearly visible (\oiii/\hbeta = 2.27, see
  Table \ref{tab:ratios}). There is evidence for the presence of the
  \oi\ $\lambda 6300$ line, although it is below the detection level
  of our spectrum. \nii\ $\lambda \lambda 6548,\,6584$ lines are
  clearly distinguished at both sides of the narrow component of
  \halpha. The broad component of \halpha\ has a FWHM of about
  3400~\kms, and its emission is comparable with that of the narrow
  component. Finally, at the red end of the spectrum the \sii\
  $\lambda \lambda 6717,\,6731$ lines can be clearly distinguished.

  The narrow components \halpha/\hbeta\ ratio has a value of 5.02,
  indicating an important amount of intrinsic reddening ($A_V =
  1.35$). The BH mass estimated from the \oiii\ FWHM is $\sim 2
  \times 10^8\ \rm{M}_{\odot}$.

  The loci of all the emission line ratios in the diagnostic
  diagrams (Fig.~\ref{fig:diagnos}) are in the AGN domain. The
  spectrum presented in this paper is consistent with that of a
  Sy~1.9, in contrast with both the SBS catalogue and the SDSS
  spectrum, which favor a Sy~1.8 classification. Of course, the
  classification may depend on the S/N ratio of the different
  spectra as well as the somewhat subjective detection threshold for
  the broad component. However, \citet{penston} have shown that
  large variations in the broad components of the Balmer lines are
  possible, and the same object may be classified as Sy~1 or Sy~2
  nuclei in different epochs. Thus, the possibility that the object
  shows some kind of variability cannot be discarded.

  About the companion, it is separated 33\arcsec\ from the SBS
  object. The systemic velocity difference between both galaxies is
  $420 \pm 30$~\kms. With respect to the permitted lines, only
  \halpha\ and \hbeta\ are observed. Although both of them are
  narrow, \hbeta\ is more than twice broader than \halpha. The
  \halpha/\hbeta\ ratio is rather large (7.40, see
  Table~\ref{tab:ratios}), indicating that intrinsic reddening is
  important ($A_V=1.35$) and the optical depth is larger than the
  thin limit ($\tau > 1$; see Table~\ref{tab:redden}). \hei\
  $\lambda 5876$ is probably present but below the level of $3
  \sigma$ detection. For the forbidden lines, the \nii\ pair is
  observed and these lines have similar FWHM to that of \halpha, but
  the \sii\ profiles are significantly broader. \oiii\ $\lambda
  5007$ is probably present but below the $3 \sigma$ detection
  level, as is \oi\ $\lambda 6300$. These oxygen lines are also very
  weak in the SDSS spectrum.

  The visual aspect (bright nucleus) and a glance to the narrow line
  spectrum suggest that the companion is a SBN. However, the line
  ratios are not well determined and two of three upper limits are
  compatible with both \hii\ regions and low activity AGN (see
  Fig.~\ref{fig:diagnos}). Only the \nii/\halpha\ upper limit is
  incompatible with an AGN. The FWHM of \hbeta\ (450 \kms) is too
  broad for a SBN or a \hii\ region, although the line is weak (it
  is barely detected at the $3\sigma$ level) and all the other lines
  have FWHM less than 300 \kms. Following \citet{oster89}, the
  \sii~$\lambda 6717$/\sii~$\lambda 6731$ line ratio (1.3) yields an
  electron density $\simeq 100\ \rm{cm}^{-3}$ (for a range of
  temperatures between 5000 and $2 \times 10^4\,\rm{K}$) which is
  typical for the spread-out giant \hii\ regions. Thus, the
  companion galaxy can be classified as a SBN.

\subsection{\protect\objectname[]{SBS 1340+569}}

  This object is a ROSAT and 2MASS source, and it is classified as a
  Sy~1.8 galaxy in the SBS catalogue. It has an emission line
  neighbor galaxy located $4\fm 4$ to the East and $14\arcsec$ to
  the South. The SDSS image for this neighbor does not show a
  star-like nucleus, and the spectra show that both objects share
  the same redshift (0.04) and that the neighbor is probably a SB
  galaxy.

  The spectrum of SBS~1340+569 does not show any \feii\ blends,
  which favors the detection of the \heii\ $\lambda 4686$ line. The
  intensity of the broad components of \halpha\ and \hbeta\ are
  approximately twice the strength of the narrow components. The
  \hbeta\ components are significantly wider than those of \halpha.
  The narrow components \halpha/\hbeta\ ratio (3.26) almost matches
  the intrinsic value accepted for AGNs (3.1), indicating a very low
  intrinsic reddening (Table~\ref{tab:redden}).

  The \hei\ $5876$ line is very weak, and the measured FWHM for this
  line (680 \kms) is probably inaccurate. The \oiii\ lines are
  clearly visible (\oiii$/$\hbeta~$\gtrsim 3$), although these lines
  are narrower than \hbeta. \oi\ $\lambda 6300$ is also present, and
  it has a FWHM similar to \oiii. The \fex\ $\lambda 6374$ is much
  broader than the other observed narrow lines (1560~\kms), as in
  the case of SBS~1118+541. The \nii\ lines are weak
  (\nii/\halpha~$=0.23$), but they can be distinguished superposed
  to the broad \halpha\ component. These lines and the narrow
  \halpha\ component have similar widths (around 110~\kms).

  The size of the BLR estimated from the continuum at 5100 \AA\ and
  the broad component of \hbeta\ luminosities are small, around 6
  light days. This feature, along with the X-ray emission detected
  by ROSAT and the small amount of reddening measured, probably
  makes of this object a good candidate for reverberation mapping
  studies.

  The presence of broad components confirms that this is an AGN. The
  \oiii/\hbeta\ line ratio ($\gtrsim 3$) indicates an ultraviolet
  excess probably associated to the AGN. The BH mass calculated from
  the \hbeta\ and \oiii\ widths agree in an estimate of
  approximately $10^7~\rm{M}_{\odot}$. Fig.~\ref{fig:critdens} shows
  the strong correlation between the FWHM and the critical density
  for each forbidden line present in SBS~1340+569. This correlation
  shows that the velocities $v$ of the clouds scale as $v \propto
  n_e^{0.23 \pm 0.01}$. This is a rather large dependency compared
  to M81 \citep{filip88} or SBS~118+541, and it may indicate the
  presence of a starburst component that contributes to the
  low-excitation and lower critical density lines. Besides, some of
  the FWHM of the narrow lines are smaller than the value of
  300~\kms\ that separates \hii\ regions from AGN, while some are
  larger than this value (Table~\ref{tab:widths}). Of the two line
  ratios plotted in the diagnostic diagrams (Fig.~\ref{fig:diagnos};
  note that the \sii\ lines are not in the range of the available
  spectrum), one lies in the AGN region while the other in the \hii\
  region. These features suggest that this object has a composite
  spectrum, and should be considered as a Seyfert (probably a
  Sy~1.8) with a SBN component.

\subsection{\protect\objectname[]{SBS 1626+554}}

  This object is an infrared (2MASS) and X-ray (ROSAT and EINSTEIN)
  source. Its spectrum clearly corresponds to a QSO, and the only
  forbidden lines observed are the \oiii\ $\lambda \lambda
  4959,\,5007$. The permitted lines do not show narrow components,
  but two broad components are necessary to fit the Balmer lines,
  whose profiles are slightly asymmetric because the broadest
  component is systematically redshifted.

  The \hgamma\ profile is complicated because it is blended with the
  \oiii\ $\lambda 4363$ line. As shown in Fig.~\ref{sbs1626hg}, the
  broadest component of \hgamma\ is redshifted by $\simeq 1000$
  \kms. A similar redshifted component can be measured in \hbeta, as
  shown in Fig.~\ref{sbs1626hb} and \halpha. The extreme of the red
  wing of the \halpha\ profile is bitten by the telluric absorption
  band between 7570 and 7700~\AA. All these lines show a fit which
  is not as good near the peak as in the rest of the line,
  suggesting the presence of a narrow component, however, too weak
  to be measured with confidence (S/N$\sim$1.5).

  Apart from the Balmer lines, the only permitted lines observed are
  \heii\ $\lambda 4686$ (after subtracting the \feii\ blends), a
  feature identified as \hei\ $\lambda 5048$, and \hei\ $\lambda
  5876$ which is easily recognized in the spectrum.

  The broad line referred as \hei\ $\lambda 5048$ deserves a
  especial remark. This feature is clearly seen as the reddest line
  in Fig.~\ref{sbs1626hb}. The \hei\ denomination is suggested
  because the measured gravity center of the line in the QSO
  rest-frame is at $\lambda 5051 \pm 3\ \AA$ and, moreover, there
  are also other \hei\ and \heii\ lines detected in the spectrum.
  However, other atomic ions have also permitted lines that dwell in
  the same region. \citet{peterson85} found a emission line at
  5050~\AA\ in Akn 120 that \citet{crenshaw} proposed to be
  \ion{Si}{2} $\lambda 5056$ o perhaps \ion{Si}{2} $\lambda 5041$.
  A similar feature was also reported by \citet{sergeev} in the
  spectrum of NGC 5548.

  An indication of the temperature and density in the NLR can be
  obtained from the \oiii\ lines. Following \citet{oster89}, the sum
  of the \oiii\ $\lambda 4959$ and \oiii\ $\lambda 5007$ fluxes
  divided by the \oiii\ $\lambda 4363$ flux has a value of 7.35,
  which indicates either a high temperature ($\lesssim 10^5\,\rm{K\
  for}\ N_e < 10^7$) or most probably a high electron density ($1.4
  \times 10^4\,\rm{K\ for}\ N_e \sim 10^7$).

  Finally, the BH mass estimated from the \hbeta\ and \oiii\ line
  widths approximately agree within a factor of 2 ($M_{BH} = 8\
  \rm{and}\ 4 \times 10^8\,\rm{M}_{\odot}$, respectively, see Table
  \ref{tab:masses}).


\section{Conclusions} \label{sec:conclusions}

  Spectroscopic observations of 8 SBS extragalactic sources, as well
  as one companion galaxy, have been made using the OAN 2.1-m
  telescope at San Pedro M{\'a}rtir, Mexico. The spectra in all but one
  case (SBS~0944+540) covers the rest frame \hbeta\ and \halpha\
  regions with a spectral resolution of 4.5~\AA. The data has been
  corrected for redshift and for galactic extinction in order to
  derive spectrophotometric parameters and line ratios.

  The spectra have been analyzed in detail and several faint
  emission lines have been measured. Any feature that strongly
  suggests the presence of an emission line was also reported.
  \feii\ blends have been subtracted and their EW measured. These
  subtractions have allowed a better estimate of the emission lines
  parameters for \hbeta\ and the \oiii\ $\lambda \lambda
  4959,\,5007$. Frequently, the \feii\ subtraction has also made
  possible the identification and measurement of \heii\ $\lambda
  4686$ and, in one case, a feature that is identified as \hei\
  $\lambda 5048$.

  Gaussian profiles were used to fit both the broad and the narrow
  emission lines. In most cases, a single component was used, except
  for the Balmer lines, which usually presented a narrow as well as
  one or more broad components. Some line profiles can also be
  described using Lorentzian fits with a lesser number of
  components. In these cases, the Lorentzian parameters for the
  broad lines have been reported.

  Emission line FWHM and line ratios have been used, along with
  diagnostic diagrams, to separate AGNs from \hii\ regions-like.
  Thus it was possible to review the classification of all objects.
  In two cases, a revised classification is proposed.

  The analysis of the spectra reveal that the widths of the lines
  frequently change along the spectral range, an effect that may
  imply a stratification of the NLR.

  Intrinsic reddening, \bv\ color excess, $A_V$ extinction
  coefficients and optical depths for the $V$ band, and at
  wavelengths corresponding to \halpha, \hbeta\ and 5100~\AA, were
  calculated for all the objects with narrow Balmer components, and
  assuming that the local reddening has a continuous distribution
  along the line of sight and that the optical properties of the
  dust in the host-galaxy are the same as in the solar neighborhood.

  Several parameters of the central black hole have been calculated
  with different methodologies and the results are compared. The
  mass of the central supermassive black hole has been inferred from
  its correlation with both the bulge stellar velocity dispersion
  (measured from the \oiii\ $\lambda 5007$ FWHM), and from the
  luminosity and FWHM of \hbeta. The luminosity has been estimated
  from both the monochromatic flux at 5100~\AA\ and the integrated
  flux of the broad \hbeta\ component.

  Below is a summary of the main topics discussed for each object:

  \begin{itemize}

       \item SBS~0848+526 is a reddened SBN which has a plume
       extending to a brighter neighbor Seyfert galaxy. The plume
       emits \halpha, \hbeta\ and \oiii\ lines.

       \item SBS~0944+540 is a QSO with strong \feii\ blends and
       without evidence of narrow lines. The \hbeta\ profile is
       strongly asymmetric and shows a large blue bump and extended
       wings.

       \item  SBS~1118+541 and 1136+595 are NLS1 which present many
       similarities. Both show discrepancies in the FWHM of the
       broad Balmer components used for their classification,
       depending on whether they were fitted using Gaussian or
       Lorentzian profiles. Their line ratios calculated from the
       Gaussian fits correspond to \hii\ regions-like. There are
       differences in the widths among all the narrow lines, and
       particularly the narrow Balmer components are much broader
       than the forbidden lines, suggesting a stratification of the
       NLR. Both objects show low amounts of reddening and the
       masses of their black holes, estimated from the Lorentzian
       fit for \hbeta, are one order of magnitude below the mass
       estimation from the Gaussian fit or from the \oiii\ FWHM.

       \item SBS~1136+594 is a Sy~1.5 galaxy showing a large amount
       of reddening. The mass of the black hole estimated from the
       \oiii\ FWHM is rather low for a Sy~1.5 and a factor 44 times
       smaller than the mass calculated from the \hbeta\ FWHM.

       \item SBS~1204+505B is an AGN with at least one companion
       galaxy, and is more probably a member of a compact group. The
       nucleus is reddened, and, in contrast with \halpha, the broad
       component of \hbeta\ is not detected in the spectrum
       presented in this paper. This feature leads to classify the
       object as a Sy~1.9, rather than a Sy~1.8 as proposed in the
       SBS catalogue. The companion of SBS~1204+505B reported in
       this paper is a very reddened SBN. However, the weak \hbeta\
       line has a FWHM of 450 \kms, which is larger than the width
       expected for this kind of objects.

       \item SBS~1340+569 is probably a Sy~1.8 galaxy with a SBN
       component. The diagnostic plots for the available line ratios
       cover both the AGN and \hii\ region domains. Besides, there
       is a variety of line widths for different ions and between
       the broad and narrow components of \halpha\ and \hbeta, being
       the widths of the narrow lines in the \halpha\ region rather
       small for an AGN. The amount of reddening is small, as well
       as the size estimated for the BLR ($\sim 6$ light days).

       \item SBS~1626+554 is a QSO whose permitted lines need two
       broad profiles to be fitted. The broadest component is
       redshifted by about 1000 \kms. Narrow components are
       suggested for all these three lines, but their S/N are too
       low to be measured reliably. The \heii\ $\lambda 4686$ is
       discernible after the subtraction of the strong \feii\
       blends. A peculiar broad emission line, probably \hei\
       $\lambda 5048$, is observed and its parameters measured.

  \end{itemize}




\acknowledgments

  The author thanks the SPM staff, in particular to Felipe Montalvo
  and Juan Gabriel Garc{\'\i}a, for their excellent assistance and
  technical support at the 2.1-m telescope. Jivan Stepanian
  suggested a list of interesting SBS objects. Luc Binette the
  anonimous referee made valuable comments that helped to improve to
  the manuscript. This work was supported by the \emph{DGAPA-UNAM}
  grant IN113002.

  The software system Image Reduction and Analysis Facility (IRAF)
  is distributed by the National Optical Astronomy Observatories,
  which are operated by the Association of Universities for Research
  in Astronomy, Inc., under cooperative agreement with the National
  Science Foundation.

  This research has made use of the NASA/IPAC Extragalactic Database
  (NED) which is operated by the Jet Propulsion Laboratory,
  California Institute of Technology, under contract with the
  National Aeronautics and Space Administration.


\clearpage



\figcaption[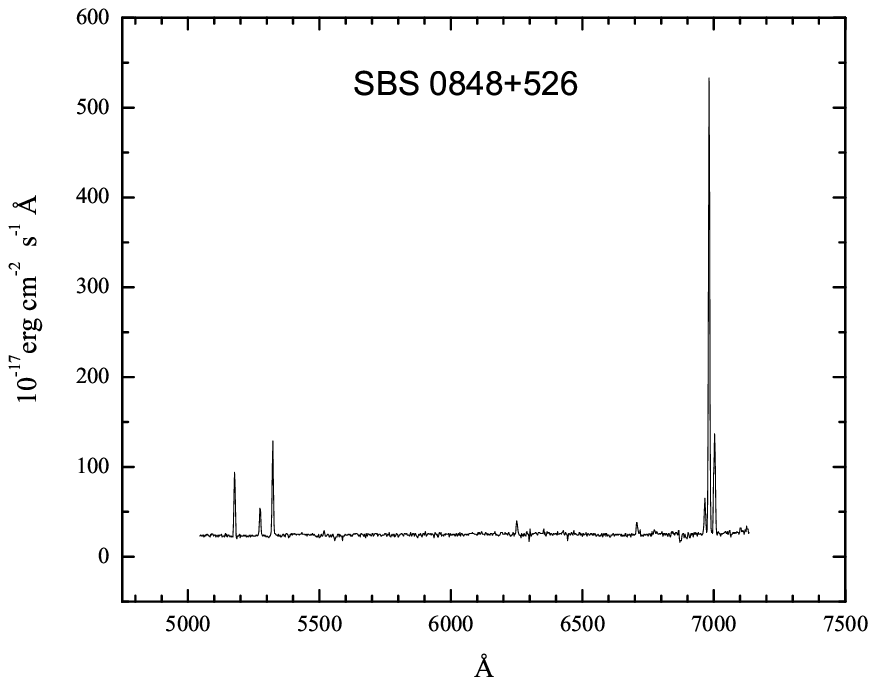,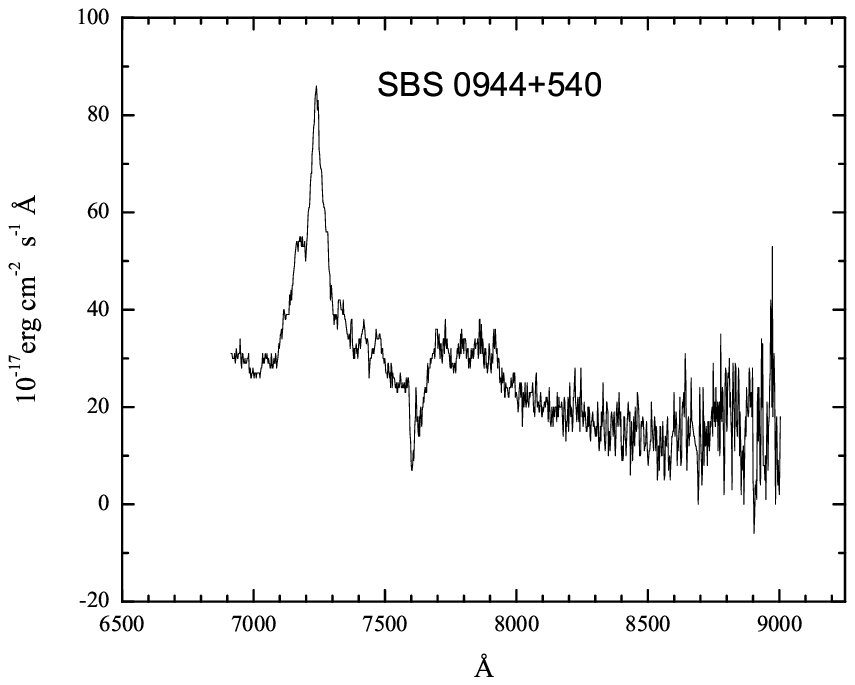,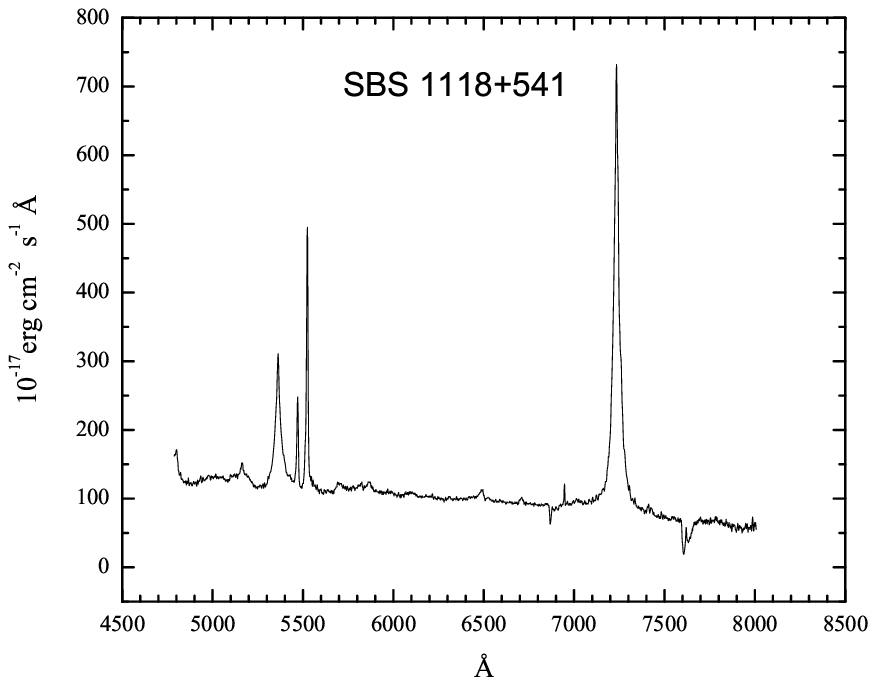,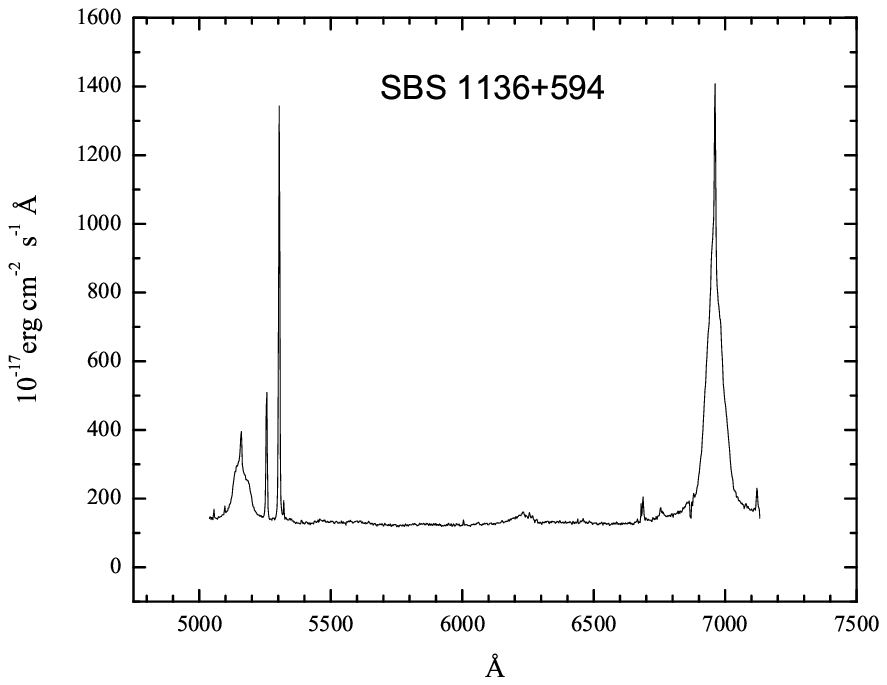,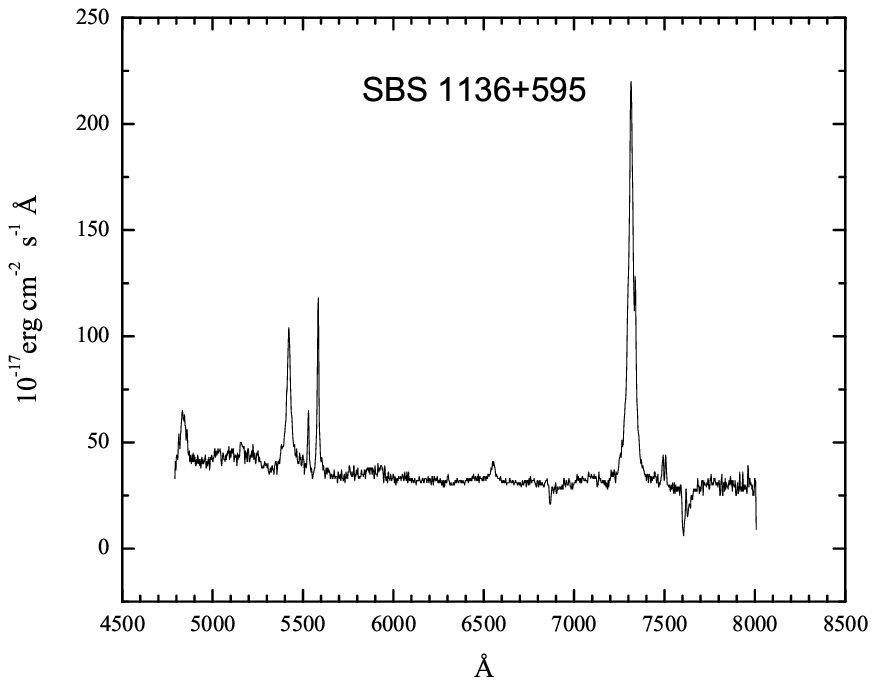, 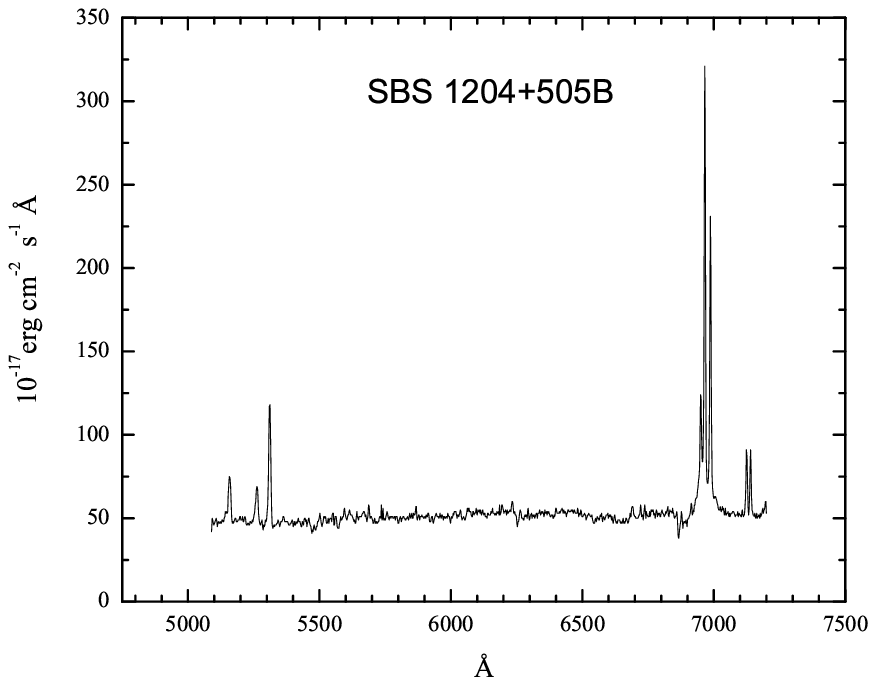,
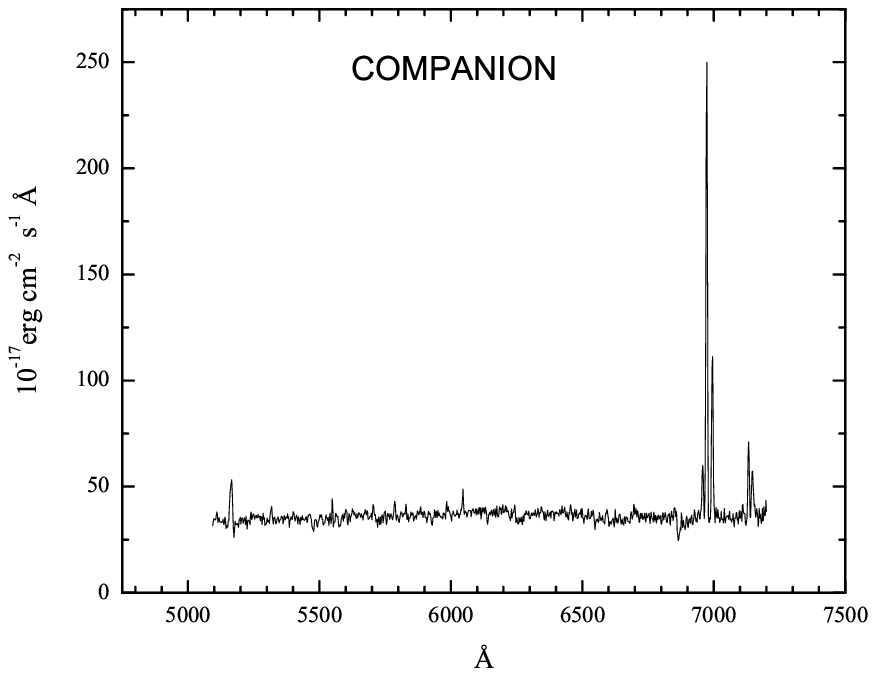, 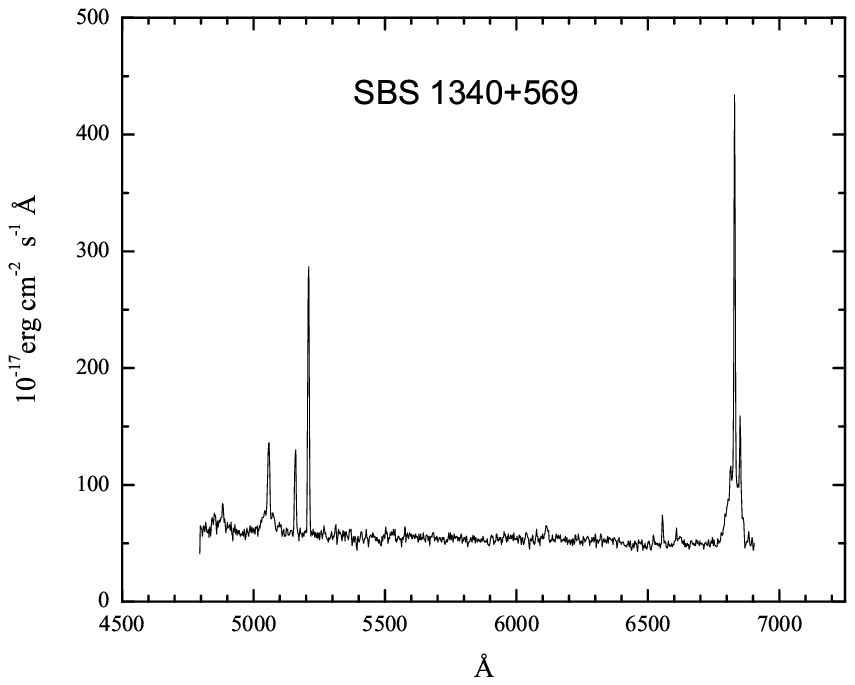, 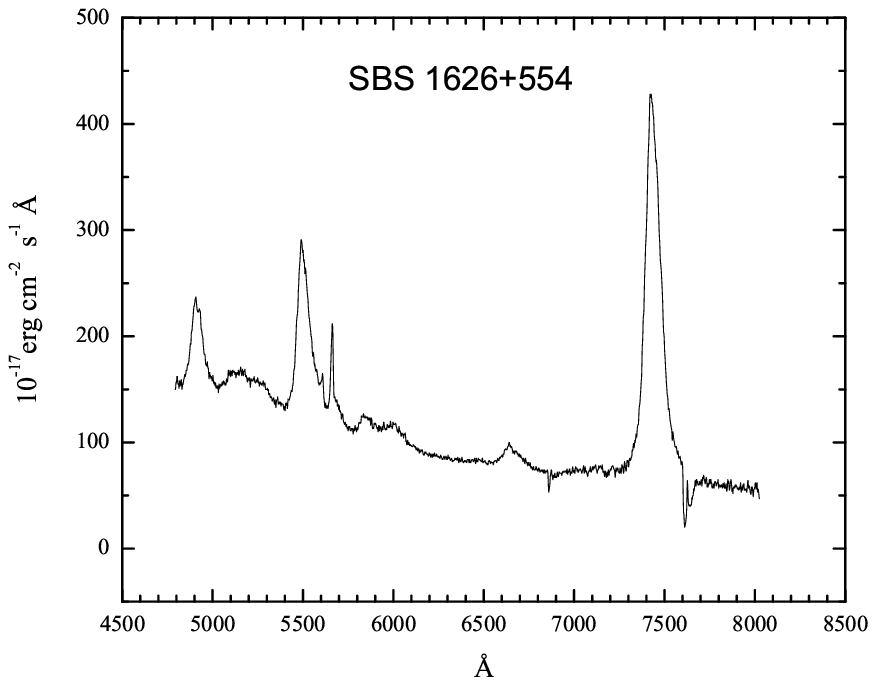]{Spectra of the SBS objects and the
companion galaxy of SBS~1204+505B. \label{fig:spec}}

\figcaption[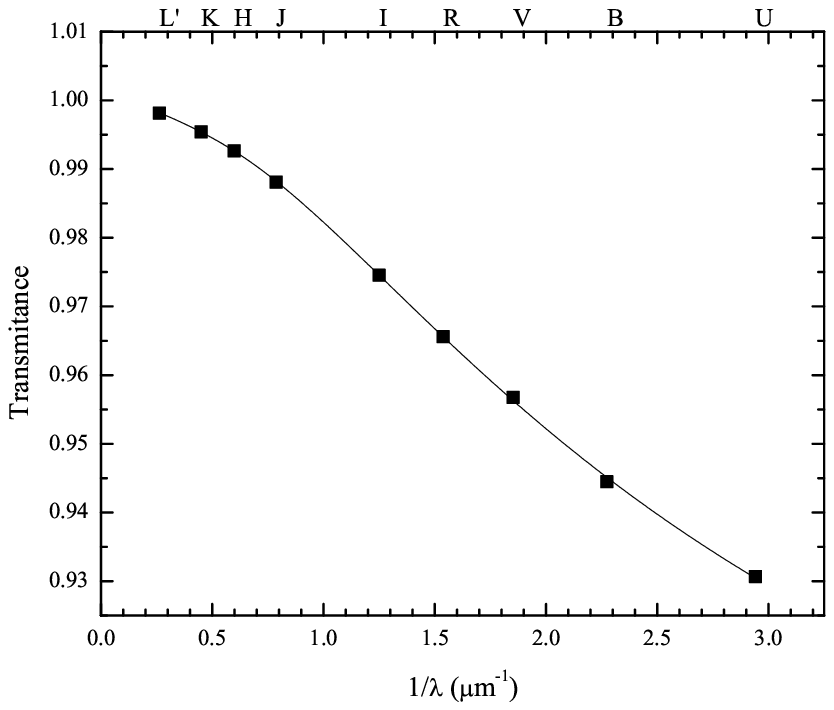]{Transmittance values for the galactic extinction
of SBS~1136+594. The squares show the transmittances obtained from
the galactic extinction (eq. \ref{eq:transmit}) for the UBVRIJHKL'
band measurements extracted from NED. The continuous line
corresponds to the exponential associate function fit (eq.
\ref{eq:exasf}). \label{fig:tranmit}}

\figcaption[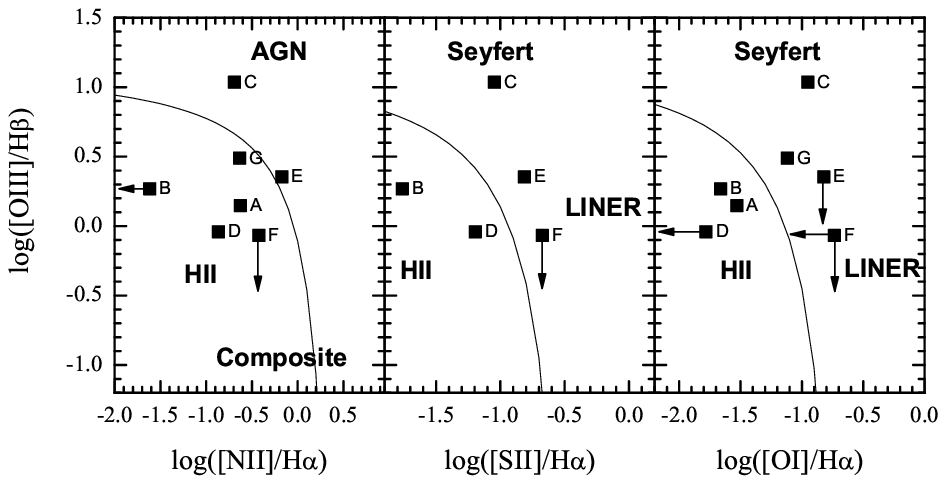]{Diagnostic diagram adapted from Kewley et al.
2006. The letter code for the objects is: A - SBS 0848+526, B - SBS
1118+541, C - SBS 1136+594, D - SBS 1136+595, E - SBS 1204+505B, F -
Companion, and G - SBS 1340+569 \label{fig:diagnos}}

\figcaption[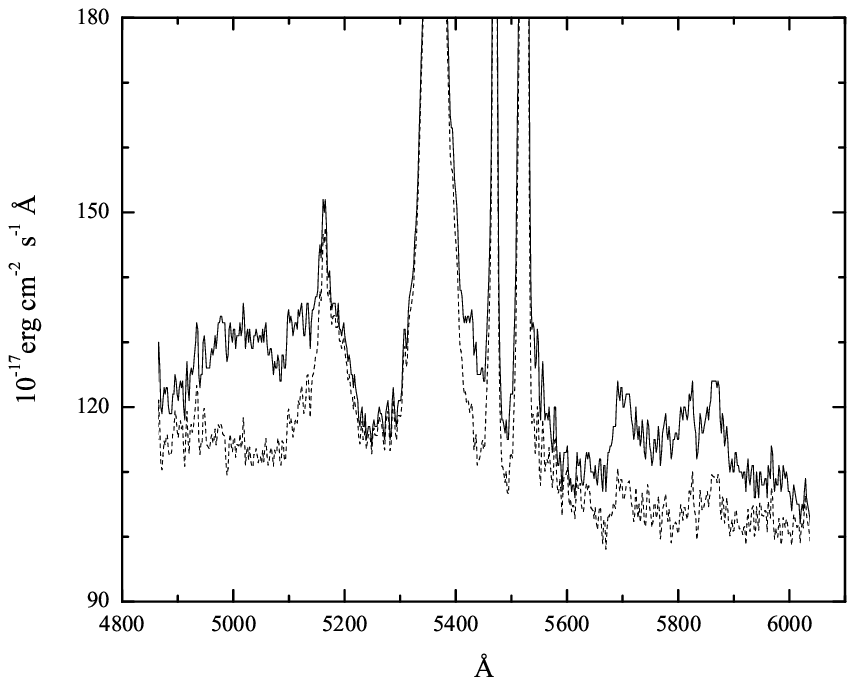]{Detail of the \hbeta\ region in SBS~1118+541.
The continuous line shows the original data, and the dashed line the
data after subtracting the \feii\ blends. The \heii\ $\lambda 4686$
(around $5170~\AA$ at the observer's frame) is clearly visible after
the subtraction. \label{fig:hb1118}}

\figcaption[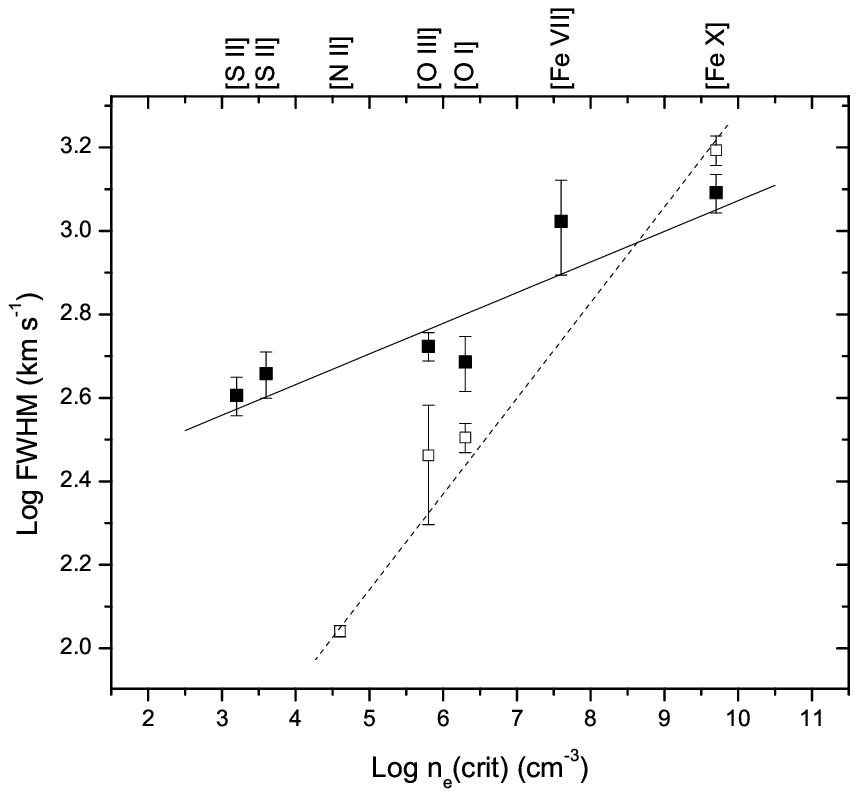]{The common logarithm of the FWHM vs the critical
density for collisional de-excitation (for a temperature of
10$^4$\,K) reveals a highly significant correlation for SBS~1118+540
(filled squares) and SBS~1340+569 (open squares). The solid line
shows the weighted linear fit for SBS~1118+540 (slope $b = 0.07 \pm
0.01$; correlation coefficient $r = 0.946$). Similarly, the dashed
line shows the weighted linear fit for SBS~1340+569 ($b = 0.23 \pm
0.01$; $r = 0.997$). \label{fig:critdens}}

\figcaption[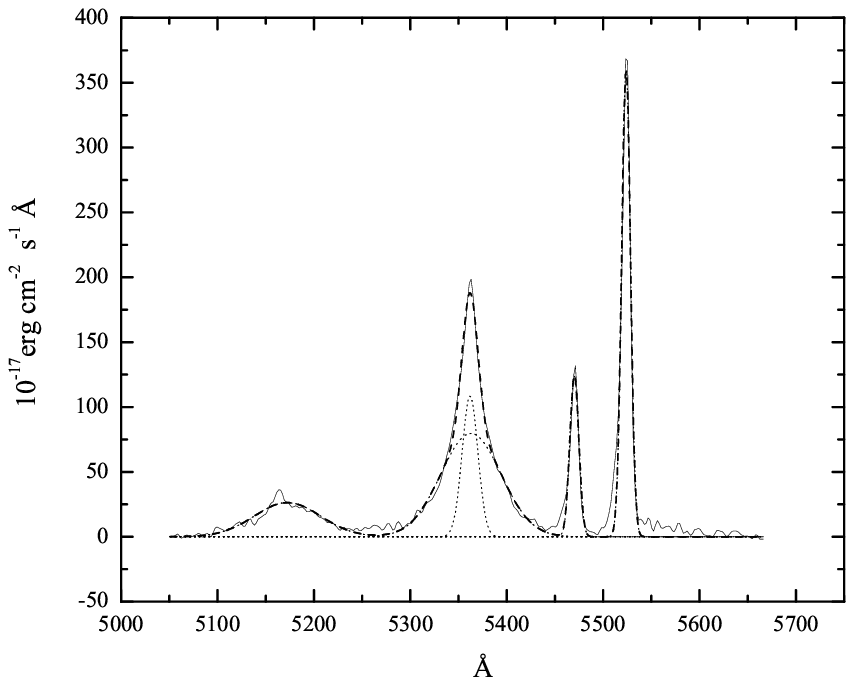,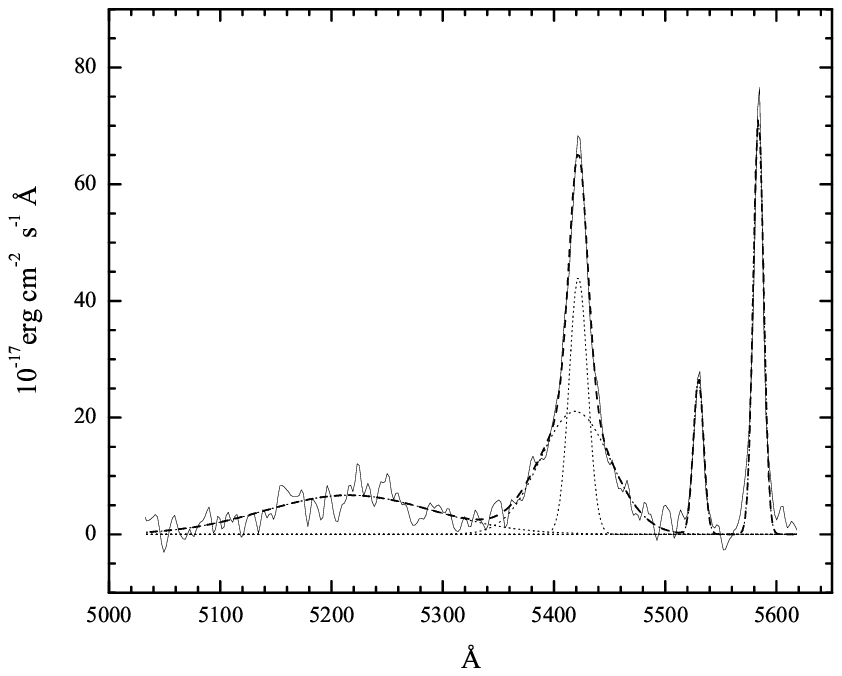]{Gaussian fit of the \hbeta\ region in
SBS~1118+541 (upper panel) and SBS~1136+595 (lower panel). The
continuous line shows the data after \feii\ and continuum
subtraction. The dotted lines show the individual components of the
fit (\heii\ $\lambda 4686$, \hbeta\ narrow and broad, and \oiii\
$\lambda 4959$ and 5007). Finally, the dashed thick line shows the
profile obtained combining the individual components.
\label{fig:hbfit}}

\figcaption[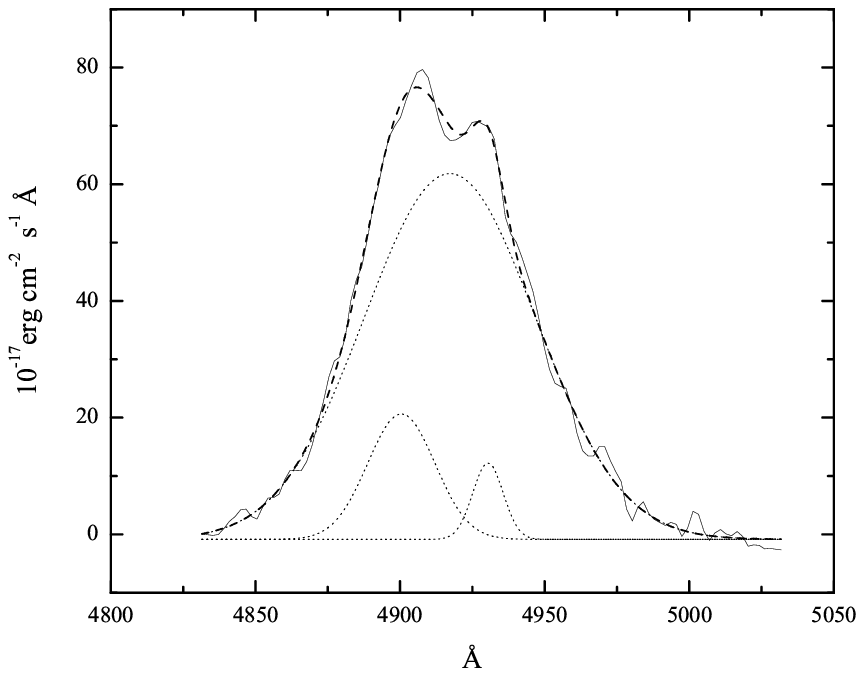]{Fit of the \hgamma\ line in SBS~1626+554. The
continuous line shows the data after subtracting the continuum. The
doted lines show the individual components of the line and the
narrow \oiii\ $\lambda 4363$, while the dashed thick line represents
the complete fit. A narrow component,would help to improve the fit
near the peak. \label{sbs1626hg}}

\figcaption[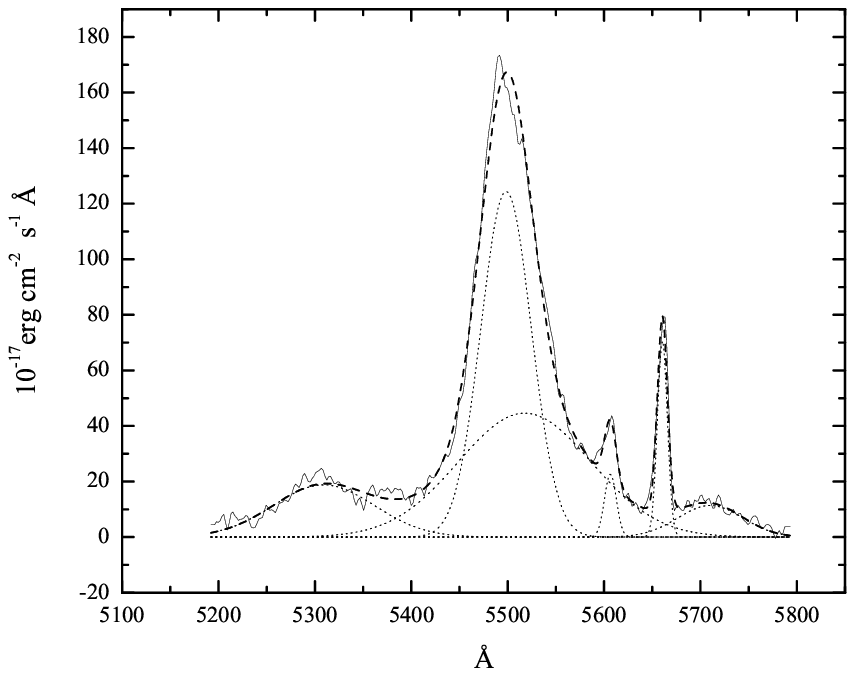]{Fit of the region around \hbeta\ in
SBS~1626+554. The continuous line shows the data after subtracting
the \feii\ blends and the continuum. The doted lines show the
individual components of the fit (\heii\ $\lambda 4686$, the three
broad components of \hbeta, and \oiii\ $\lambda \lambda
4959,\,5007$). The dashed thick line shows the complete fit. As in
the previous figure, a weak narrow component would improve the fit
near the peak. \label{sbs1626hb}}

\clearpage


\begin{deluxetable}{lllllllll}

 \tabletypesize{\footnotesize}
 \tablecaption{Log of observations \label{tab:log}}
 \tablewidth{0pt}
  \tablehead{
  \multicolumn{1}{c}{SBS} & \multicolumn{1}{c}{R.A.}
  & \multicolumn{1}{c}{Dec.} &
  & \multicolumn{1}{c}{$z$}
  &  & \multicolumn{1}{c}{Date}
  & \multicolumn{1}{c}{Spectral} & \multicolumn{1}{c}{Spectral} \\
  \multicolumn{1}{c}{designation} & \multicolumn{1}{c}{2000}
  & \multicolumn{1}{c}{2000}
  & \multicolumn{1}{c}{\raisebox{1.5ex}[0cm][0cm]{$B$}}
  & \multicolumn{1}{c}{em}
  & \multicolumn{1}{c}{\raisebox{1.5ex}[0cm][0cm]{$M_B$}}
  & \multicolumn{1}{c}{dd-mm-yy}
  & \multicolumn{1}{c}{range} & \multicolumn{1}{c}{type\tablenotemark{a}}}

  \startdata
    0848+526\tablenotemark{b}
                 & 08 51 51.40 & +52 28 25.2 & 17.5  & 0.064 & -19.9 & 11-04-02 & 5000-7100 & SBN        \\
    0944+540     & 09 47 55.94 & +53 50 00.5 & 16.84 & 0.488 & -25.5 & 14-04-02 & 6800-9000 & QSO        \\
    1118+541     & 11 21 08.55 & +53 51 20.7 & 16.41 & 0.103 & -22.9 & 12-04-02 & 4800-7550 & NLS1       \\
                 &             &             &       &       &       & 14-04-02 & 6500-8000 &            \\
    1136+594     & 11 39 08.91 & +59 11 54.5 & 16.26 & 0.060 & -21.5 & 11-04-02 & 5000-7100 & Sy~1.5      \\
    1136+595     & 11 39 00.48 & +59 13 46.7 & 17.0  & 0.115 & -22.1 & 12-04-02 & 4800-7580 & NLS1       \\
                 &             &             &       &       &       & 14-04-02 & 6500-8000 &            \\
    1204+505B\tablenotemark{b}
                 & 12 06 55.55 & +50 17 36.8 & 16.64\tablenotemark{c}
                                                     & 0.062 & -21.3 & 13-04-02 & 5160-7210 & \textit{Sy~1.9} \\
    (\& Companion)\tablenotemark{d}
                 & 12 06 52.40 & +50 17 24.3 & 17.08\tablenotemark{c}
                                                     & 0.063 & -20.8 & 13-04-02 & 5160-7210 & \textit{SBN}       \\
    1340+569\tablenotemark{b}
                 & 13 42 10.23 & +56 42 11.6 & 16.98 & 0.040 & -20.4 & 14-04-02 & 4800-7210 & \textit{Sy~1+SBN}    \\
    1626+554     & 16 27 56.11 & +55 22 31.9 & 16.17 & 0.134 & -23.4 & 12-04-02 & 4800-7580 & QSO        \\
                 &             &             &       &       &       & 14-04-02 & 6500-8000 &            \\

  \enddata

  \tablenotetext{a}{New or revised classification in italics.}

  \tablenotetext{b}{Pair.}

  \tablenotetext{c}{Calculated from SDSS photometric database using
  the \citet{smith} set of transformation equations.}

  \tablenotetext{d}{A companion of SBS~1204+505B.}

\end{deluxetable}


\clearpage





\begin{deluxetable}{lrrrrrrrrrrrrrrrrrr}
\setlength{\tabcolsep}{0.025in}
 \tabletypesize{\scriptsize}
 \tablecaption{Gaussian rest-frame EW and FWHM \label{tab:widths}}
 \tablewidth{0pt}
 \tablehead{
\multicolumn{1}{c}{Line}       & \multicolumn{2}{c}{0848+526} &
\multicolumn{2}{c}{0944+540}   & \multicolumn{2}{c}{1118+541} &
\multicolumn{2}{c}{1136+594}   & \multicolumn{2}{c}{1136+595} &
\multicolumn{2}{c}{1204+505B}  & \multicolumn{2}{c}{Companion}   &
\multicolumn{2}{c}{1340+569}   & \multicolumn{2}{c}{1626+554}
\\
\cline{2-19} \noalign{\smallskip}
   \multicolumn{1}{c}{identification}
 & \multicolumn{1}{c}{EW} & \multicolumn{1}{c}{FWHM}
 & \multicolumn{1}{c}{EW} & \multicolumn{1}{c}{FWHM}
 & \multicolumn{1}{c}{EW} & \multicolumn{1}{c}{FWHM}
 & \multicolumn{1}{c}{EW} & \multicolumn{1}{c}{FWHM}
 & \multicolumn{1}{c}{EW} & \multicolumn{1}{c}{FWHM}
 & \multicolumn{1}{c}{EW} & \multicolumn{1}{c}{FWHM}
 & \multicolumn{1}{c}{EW} & \multicolumn{1}{c}{FWHM}
 & \multicolumn{1}{c}{EW} & \multicolumn{1}{c}{FWHM}
 & \multicolumn{1}{c}{EW} & \multicolumn{1}{c}{FWHM}
}
\startdata
%
\hgamma~$\lambda 4340$  & \nd & \nd & \nd &  \nd &\nd &  \nd & \nd &  \nd & 22 & 2420 &\nd & \nd &\nd &  \nd &\nd &  \nd &  31 &  4250 \\
~~2nd component         & \nd & \nd & \nd &  \nd &\nd &  \nd & \nd &  \nd &\nd &  \nd &\nd & \nd &\nd &  \nd &\nd &  \nd &   4 &  1650 \\
\oiii~$\lambda 4363$    & \nd & \nd & \nd &  \nd &\nd &  \nd & \nd &  \nd &\nd &  \nd &\nd & \nd &\nd &  \nd &\nd &  \nd &   1 &   710 \\
\feii~$\lambda 4540$    & \nd & \nd & \nd &  \nd & 29 &  \nd & \nd &  \nd & 37 &  \nd &\nd & \nd &\nd &  \nd &\nd &  \nd &  54 &   \nd \\
\heii~$\lambda 4686$    & \nd & \nd & $<2$&  \nd & 19 & 4690 & \nd &  \nd & 33 &10240 &\nd & \nd &\nd &  \nd &  9 & 3560 &  13 &  5650 \\
\\
\hbeta~$\lambda 4861$n  &  17 & 210 & \nd &  \nd & 19 & 1040 &   4 &  250 & 24 & 1050 &  6 & 520 &  5 &  450 &  9 &  400 & \nd &   \nd \\
\hbeta~$\lambda 4861$b  & \nd & \nd &  33 & 2630 & 53 & 4110 &  61 & 2920 & 44 & 4120 &\nd & \nd &\nd &  \nd & 18 & 3430 &  61 &  8860 \\
~~2nd component         & \nd & \nd &  81 & 2470 &\nd &  \nd &  13 & 1720 &\nd &  \nd &\nd & \nd &\nd &  \nd &\nd &  \nd &  65 &  3350 \\
~~3rd component         & \nd & \nd & 112 &13120 &\nd &  \nd & \nd &  \nd &\nd &  \nd &\nd & \nd &\nd &  \nd &\nd &  \nd & \nd &   \nd \\
\\
\oiii~$\lambda 4959$    &   8 & 200 & \nd &  \nd & 12 &  550 &  17 &  220 &  8 &  530 &  5 & 510 &\nd &  \nd &  9 &  330 &   3 &   620 \\
\oiii~$\lambda 5007$    &  23 & 200 & \nd &  \nd & 36 &  530 &  51 &  200 & 23 &  540 & 13 & 500 &$<5$&  \nd & 28 &  290 &   8 &   610 \\
\hei\ $\lambda 5048$    & \nd & \nd & \nd &  \nd &\nd &  \nd & \nd &  \nd &\nd &  \nd &\nd & \nd &\nd &  \nd &\nd &  \nd &   6 &  4340 \\
\feii~$\lambda 5300$    & \nd & \nd & 101 &  \nd & 46 &  \nd &  34 &  \nd & 43 &  \nd &\nd & \nd &\nd &  \nd &\nd &  \nd &  78 &   \nd \\
\\
\hei~$\lambda 5876$     &   4 & 170 & \nd &  \nd &  8 & 5350 &  11 & 3990 &  6 & 1210 &\nd & \nd &$<5$&  \nd &  3 &  680 &  30 &  5460 \\
~~2nd component         & \nd & \nd & \nd &  \nd &  2 &  920 & \nd &  \nd &\nd &  \nd &\nd & \nd &\nd &  \nd &\nd &  \nd & \nd &   \nd \\
\fevii~$\lambda 6087$   & \nd & \nd & \nd &  \nd &  2 & 1050 & $<4$&  \nd &\nd &  \nd &\nd & \nd &\nd &  \nd &\nd &  \nd & \nd &   \nd \\
\oi~$\lambda 6300$      &   3 & 160 & \nd &  \nd &  1 &  490 &   4 &  340 &\nd &  \nd &$<5$&     &$<7$&  \nd &  3 &  320 & \nd &   \nd \\
\fex~$\lambda 6374$     & \nd & \nd & \nd &  \nd &  2 & 1230 &   3 &  330 &\nd &  \nd &\nd & \nd &\nd &  \nd &  6 & 1560 & \nd &   \nd \\
\\
\nii~$\lambda 6548$     &   9 & 150 & \nd &  \nd &\nd &  \nd &   2 &  320 &  7 &  350 &  7 & 200 &  5 &  200 &  3 &  120 & \nd &   \nd \\
\halpha~$\lambda 6563$n & 116 & 150 & \nd &  \nd & 90 &  870 &  32 &  320 &140 & 1000 & 30 & 200 & 39 &  200 & 38 &  110 & \nd &   \nd \\
\halpha~$\lambda 6563$b & \nd & \nd & \nd &  \nd &220 & 2970 & 222 & 3210 &176 & 4870 & 37 &3390 &\nd &  \nd & 80 & 2260 & 320 &  3220 \\
~~2nd component         & \nd & \nd & \nd &  \nd &\nd &  \nd & 126 & 3600 &\nd &  \nd &\nd & \nd &\nd &  \nd &\nd &  \nd & 289 &  6680 \\
\nii~$\lambda6584$      &  27 & 190 & \nd &  \nd &\nd &  \nd &   6 &  320 & 19 &  350 & 20 & 200 & 15 &  210 &  9 &  110 & \nd &   \nd \\
\\
\sii~$\lambda6717$      & \nd & \nd & \nd &  \nd &  1 &  400 &   3 &  320 &  8 &  510 &  4 & 200 &  6 &  280 &\nd &  \nd & \nd &   \nd \\
\sii~$\lambda6731$      & \nd & \nd & \nd &  \nd &  1 &  460 & \nd & \nd  &  7 &  460 &  4 & 200 &  5 &  290 &\nd &  \nd & \nd &   \nd \\
\enddata
%
\end{deluxetable}


\clearpage




\begin{deluxetable}{lrrrrrr}
 \tablecaption{Lorentzian rest-frame EW and FWHM \label{tab:lore}}
 \tablewidth{0pt}
  \tablehead{
  \multicolumn{1}{c}{Line} & \multicolumn{2}{c}{0944+540}   &
  \multicolumn{2}{c}{1118+541} & \multicolumn{2}{c}{1136+595}
  \\
  \cline{2-7} \noalign{\smallskip}
     \multicolumn{1}{c}{identification}
   & \multicolumn{1}{c}{EW} & \multicolumn{1}{c}{FWHM}
   & \multicolumn{1}{c}{EW} & \multicolumn{1}{c}{FWHM}
   & \multicolumn{1}{c}{EW} & \multicolumn{1}{c}{FWHM}
  }
  \startdata
  \hgamma~$\lambda 4340$  & \nd &  \nd &\nd &  \nd & 28 & 1930 \\

  \heii~$\lambda 4686$    & $<2$&  \nd & 22 & 3270 & 42 & 8700 \\

  \hbeta~$\lambda 4861$b  & 157 & 2700 & 73 & 1510 & 73 & 1480 \\

  ~~2nd component         &  65 & 3220 &\nd &  \nd &\nd &  \nd \\

  \hei~$\lambda 5876$     & \nd &  \nd & 24 & 5220 & 13 & 1360 \\

  \halpha~$\lambda 6563$b & \nd &  \nd &355 & 1380 &334 & 1340 \\
  \enddata
\end{deluxetable}


\clearpage


\begin{deluxetable}{lrrrrrr}
  \tablecaption{Gaussian narrow emission line ratios \label{tab:ratios}}
   \tablewidth{0pt}
  \tablehead{
    \multicolumn{1}{c}{SBS}
    & \multicolumn{1}{c}{\underline{\oiii $\lambda 5007$}}
    & \multicolumn{1}{c}{\underline{\feii $\lambda 5300$}}
    & \multicolumn{1}{c}{\underline{\oi $\lambda 6300$}}
    & \multicolumn{1}{c}{\underline{\halpha n}}
    & \multicolumn{1}{c}{\underline{\nii $\lambda 6584$}}
    & \multicolumn{1}{c}{\underline{\sii $\lambda 6717$}}
  \\
    \multicolumn{1}{c}{Designation}
    & \multicolumn{1}{c}{\hbeta n}
    & \multicolumn{1}{c}{\hbeta n}
    & \multicolumn{1}{c}{\halpha n}
    & \multicolumn{1}{c}{\hbeta n}
    & \multicolumn{1}{c}{\halpha n}
    & \multicolumn{1}{c}{\halpha n}
  }

  \startdata


  0848+526  &  1.41 &  \nd &  0.03 & 6.84 &  0.24 &  \nd \\

  1118+541  &  1.86 & 1.81 &  0.02 & 3.60 &$<0.03$& 0.02 \\

  1136+594  & 10.88 & 6.53 &  0.11 & 7.83 &  0.20 & 0.09 \\

  1136+595  &  0.91 & 1.61 &$<0.02$& 4.24 &  0.14 & 0.06 \\

  1204+505B &  2.27 &  \nd &$<0.15$& 5.02 &  0.67 & 0.15 \\

  Companion &$<0.86$&  \nd &$<0.18$& 7.40 &  0.38 & 0.21 \\

  1340+569  &  3.09 &  \nd &  0.08 & 3.26 &  0.23 &  \nd \\

  \noalign{\smallskip} \hline

  \enddata

\end{deluxetable}


\clearpage


\begin{deluxetable}{lrrrrrrr}
  \tablecaption{Reddening parameters \label{tab:redden}}
  \tablewidth{0pt}
  \tablehead{
  \multicolumn{1}{c}{SBS} &
  \multicolumn{1}{c}{$C$} &
  \multicolumn{1}{c}{$E_{B-V}$} &
  \multicolumn{1}{c}{$A_{V}$} &
  \multicolumn{1}{c}{$\tau_{V}$} &
  \multicolumn{1}{c}{$\tau_{\alpha}$} &
  \multicolumn{1}{c}{$\tau_{\beta}$} &
  \multicolumn{1}{c}{$\tau_{5100}$}
  }
  \startdata
     0848+526  & 2.21 & 0.73 & 2.22 & 2.04 & 1.66 & 2.42 & 2.26 \\
     1118+541  & 0.42 & 0.14 & 0.42 & 0.39 & 0.31 & 0.46 & 0.43 \\
     1136+594  & 2.58 & 0.85 & 2.60 & 2.39 & 1.94 & 2.83 & 2.64 \\
     1136+595  & 0.87 & 0.29 & 0.88 & 0.81 & 0.65 & 0.96 & 0.89 \\
     1204+505B & 1.34 & 0.44 & 1.35 & 1.25 & 1.01 & 1.47 & 1.38 \\
     Companion & 2.42 & 0.80 & 2.44 & 2.25 & 1.82 & 2.66 & 2.48 \\
     1340+569  & 0.14 & 0.05 & 0.14 & 0.13 & 0.10 & 0.15 & 0.14 \\
  \enddata
\end{deluxetable}

\clearpage


\begin{deluxetable}{lrrrrrrccp{0.15cm}@{}lp{0.3cm}r@{}l}
  \tabletypesize{\small}
  \tablecaption{Parameters of the black holes \label{tab:masses}}
  \tablewidth{0pt}
  \tablehead{
  &
  \multicolumn{2}{c}{$\log (L/\rm{L}_\odot)$} & &
  \multicolumn{2}{c}{$R_{\rm{BLR}}$ (light days)} & &
  \multicolumn{4}{c}{$\log (M_{\rm{BH}}/\rm{M}_\odot)$} &
  & \multicolumn{2}{c}{$\underline{M_{\rm{BH}(5100\ \AA)}}$}
  \\
  \cline{2-3} \cline{5-6} \cline{8-11}    \noalign{\smallskip}
  \multicolumn{1}{c}{\raisebox{2.0ex}[0cm][0cm]{SBS}}& 5100 \AA & \hbeta &
  & 5100 \AA & \hbeta &
  & 5100 \AA & \hbeta & \oiii &
  & \multicolumn{3}{c}{\raisebox{3.5ex}[0cm][0cm]{$\log\,\Biggl($ } \raisebox{1.5ex}[0cm][0cm]{$M_{\rm{BH}(\textrm{[\ion{O}{3})]}}$}} \raisebox{3.5ex}[0cm][0cm]{$\Biggr)$}
  }
  \startdata
     0944+540\tablenotemark{a}
               & 11.69 & 10.16 & & 169.62 & 281.19 & & 9.63 & 9.85 &\nd& \nd & &\nd & \nd \\
     1118+541  & 11.07 &  9.13 & &  63.21 &  54.54 & & 8.19 & 8.13 & 8 & .34 & & -0 & .15 \\
     1136+594  & 11.62 &  9.84 & & 151.18 & 166.90 & & 8.28 & 8.32 & 6 & .64 & &  1 & .64 \\
     1136+595  & 10.87 &  8.88 & &  45.93 &  36.30 & & 8.06 & 7.96 & 8 & .37 & & -0 & .31 \\
     1204+505B & 10.70 &  6.82 & &  35.23 &    \nd & &  \nd &  \nd & 8 & .24 & &\nd & \nd \\
     1340+569  &  9.82 &  7.38 & &   8.60 &   3.40 & & 7.17 & 6.77 & 7 & .29 & & -0 & .12 \\
     1626+554\tablenotemark{a}
               & 11.13 &  9.25 & &  69.35 &  66.09 & & 8.90 & 8.88 & 8 & .58 & &  0 & .32 \\
  \enddata
  \tablenotetext{a}{Not corrected for intrinsic reddening.}
\end{deluxetable}

\clearpage

\setcounter{figure}{0}  


   \begin{figure}
   \centering
   \includegraphics[width=0.32\linewidth,clip]{f1a.eps}
   \hfill
   \includegraphics[width=0.32\linewidth,clip]{f1b.eps}
   \hfill
   \includegraphics[width=0.32\linewidth,clip]{f1c.eps}\\
   \includegraphics[width=0.32\linewidth,clip]{f1d.eps}
   \hfill
   \includegraphics[width=0.32\linewidth,clip]{f1e.eps}
   \hfill
   \includegraphics[width=0.32\linewidth,clip]{f1f.eps}\\
   \includegraphics[width=0.32\linewidth,clip]{f1g.eps}
   \hfill
   \includegraphics[width=0.32\linewidth,clip]{f1h.eps}
   \hfill
   \includegraphics[width=0.32\linewidth,clip]{f1i.eps}\\
      \caption{
              }
   \end{figure}


\clearpage


   \begin{figure}
   \centering
   \includegraphics[width=\linewidth,clip]{f2.eps}
      \caption{
              }
   \end{figure}


\clearpage


   \begin{figure}
   \centering
   \includegraphics[width=\linewidth,clip]{f3.eps}
      \caption{
              }
   \end{figure}


\clearpage


   \begin{figure}
   \centering
   \includegraphics[width=\linewidth,clip]{f4.eps}
      \caption{
              }
   \end{figure}


\clearpage


   \begin{figure}
   \centering
   \includegraphics[width=\linewidth,clip]{f5.eps}
      \caption{
              }
   \end{figure}


\clearpage


   \begin{figure}
   \centering
   \includegraphics[width=0.7\linewidth,clip]{f6a.eps}
   \includegraphics[width=0.7\linewidth,clip]{f6b.eps}
      \caption{
              }
   \end{figure}


\clearpage


   \begin{figure}
   \centering
   \includegraphics[width=\linewidth,clip]{f7.eps}
      \caption{
              }
   \end{figure}


   \begin{figure}
   \centering
   \includegraphics[width=\linewidth,clip]{f8.eps}
      \caption{
              }
   \end{figure}


\clearpage


\begin{thebibliography}{}

    \bibitem[Antonucci(1993)]{antonuc93} Antonucci, R.\ 1993, \araa,
    31, 473

    \bibitem[Antonucci \& Miller(1985)]{antonuc85} Antonucci,
    R.~R.~J., \& Miller, J.~S.\ 1985, \apj, 297, 621

    \bibitem[Baldwin et al.(1981)]{baldwin81} Baldwin, J.~A.,
    Phillips, M.~M., \& Terlevich, R.\ 1981, \pasp, 93, 5

    \bibitem[Baldwin et al.(1995)]{baldwin95} Baldwin, J., Ferland,
    G., Korista, K., \& Verner, D.\ 1995, \apjl, 455, L119

    \bibitem[Balzano(1983)]{balzano} Balzano, V.~A.\ 1983,
    \apj, 268, 602

    \bibitem[Bian \& Zhao(2004)]{bian04} Bian, W., \& Zhao, Y.\ 2004,
    \mnras, 347, 607

    \bibitem[Bian et al.(2006a)]{bian06a} Bian, W.-H., Cui, Q.-L., \&
    Chao, L.-H.\ 2006a, Chin. J. Astron. Astrophy., 6, 281

    \bibitem[Bian et al.(2006b)]{bian06b} Bian, W., Yuan, Q.,
    \& Zhao, Y.\ 2006b, \mnras, 367, 860

    \bibitem[Binette et al.(1993)]{binette} Binette, L., Wang, J.,
    Villar-Martin, M., Martin, P.~G., \& Magris C., G.\ 1993, \apj,
    414, 535

    \bibitem[Bischof \& Becker(1997)]{bischof} Bischof, O.~B., \&
    Becker, R.~H.\ 1997, \aj, 113, 2000

    \bibitem[Boller et al.(1996)]{boller} Boller, T., Brandt, W.~N.,
    \& Fink, H.\ 1996, \aap, 305, 53

    \bibitem[Botte et al.(2005)]{botte} Botte, V., Ciroi, S., di
    Mille, F., Rafanelli, P., \& Romano, A.\ 2005, \mnras, 356,

    \bibitem[Blumenthal \& Mathews(1975)]{blumenthal} Blumenthal,
    G.~R., \& Mathews, W.~G.\ 1975, \apj, 198, 517

    \bibitem[Cardelli et al.(1989)]{cardelli} Cardelli, J.~A.,
    Clayton, G.~C., \& Mathis, J.~S.\ 1989, \apj, 345, 245

    \bibitem[Crenshaw \& Peterson(1986)]{crenshaw} Crenshaw, D.~M.,
    \& Peterson, B.~M.\ 1986, \pasp, 98, 185

    \bibitem[de Robertis \& Osterbrock(1984)]{derobert} de Robertis,
    M.~M., \& Osterbrock, D.~E.\ 1984, \apj, 286, 171

    \bibitem[Dietrich et al.(2005)]{dietrich} Dietrich, M., Crenshaw,
    D.~M., \& Kraemer, S.~B.\ 2005, \apj, 623, 700

    \bibitem[Dumont \& Collin-Souffrin(1990a)]{dumont90a} Dumont,
    A.~M., \& Collin-Souffrin, S.\ 1990a, \aaps, 83, 71

    \bibitem[Dumont \& Collin-Souffrin(1990b)]{dumont90b} Dumont,
    A.~M., \& Collin-Souffrin, S.\ 1990b, \aap, 229, 313

    \bibitem[Espey et al.(1994)]{espey} Espey, B.~R., et al.\ 1994,
    \apj, 434, 484

    \bibitem[Ferguson et al.(1997)]{ferguson} Ferguson, J.~W.,
    Korista, K.~T., Baldwin, J.~A., \& Ferland, G.~J.\ 1997, \apj,
    487, 122

    \bibitem[Ferrarese \& Merritt(2000)]{ferra00} Ferrarese, L., \&
    Merritt, D.\ 2000, \apjl, 539, L9

    \bibitem[Ferrarese et al.(2001)]{ferra01} Ferrarese, L., Pogge,
    R.~W., Peterson, B.~M., Merritt, D., Wandel, A., \& Joseph,
    C.~L.\ 2001, \apjl, 555, L79

    \bibitem[Filippenko \& Halpern(1984)]{filip84} Filippenko, A.~V.,
    \& Halpern, J.~P.\ 1984, \apj, 285, 458

    \bibitem[Filippenko \& Sargent(1985)]{filip85} Filippenko, A.~V.,
    \& Sargent, W.~L.~W.\ 1985, \apjs, 57, 503

    \bibitem[Filippenko \& Sargent(1988)]{filip88} Filippenko, A.~V.,
    \& Sargent, W.~L.~W.\ 1988, \apj, 324, 134

    \bibitem[Gaskell(1982)]{gaskell82} Gaskell, C.~M.\ 1982, \pasp,
    94, 891

    \bibitem[Gaskell \& Ferland(1984)]{gaskell84} Gaskell, C.~M., \&
    Ferland, G.~J.\ 1984, \pasp, 96, 393

    \bibitem[Gebhardt et al.(2000)]{gebhardt} Gebhardt, K., et al.\
    2000, \apjl, 539, L13

    \bibitem[Gon\c{c}alves et al.(1999)]{goncalves} Gon\c{c}alves, A.~C.,
    V{\'e}ron-Cetty, M.-P., \& V{\'e}ron, P.\ 1999, \aaps, 135,
    437

    \bibitem[Goodrich(1989)]{good} Goodrich, R.~W.\ 1989,
    \apj, 342, 224


    \bibitem[Halpern \& Steiner(1983)]{halpern} Halpern, J.~P., \&
    Steiner, J.~E.\ 1983, \apjl, 269, L37

    \bibitem[Heckman(1980)]{heckman} Heckman, T.~M.\ 1980, \aap,
    87, 152

    \bibitem[Hewitt \& Burbidge(1989)]{hewitt} Hewitt, A., \&
    Burbidge, G.\ 1989, A New Optical Catalog of Quasi-Stellar
    Objects, Univ. of Chicago Press, Chicago

    \bibitem[Ho et al.(1997)]{ho} Ho, L.~C., Filippenko, A.~V.,
    Sargent, W.~L.~W., \& Peng, C.~Y.\ 1997, \apjs, 112, 391

    \bibitem[Jim{\'e}nez-Bail{\'o}n et al.(2005)]{jim05} Jim{\'e}nez-Bail{\'o}n, E.,
    Santos-Lle{\'o}, M., Dahlem, M., Ehle, M., Mas-Hesse, J.~M.,
    Guainazzi, M., Heckman, T.~M., \& Weaver, K.~A.\ 2005, \aap, 442,
    861

    \bibitem[Jim{\'e}nez-Bail{\'o}n et al.(2003)]{jim03} Jim{\'e}nez-Bail{\'o}n, E.,
    Santos-Lle{\'o}, M., Mas-Hesse, J.~M., Guainazzi, M., Colina, L.,
    Cervi{\~n}o, M., \& Gonz{\'a}lez Delgado, R.~M.\ 2003, \apj, 593, 127

    \bibitem[Kaspi et al.(2000)]{kaspi00} Kaspi, S., Smith, P.~S.,
    Netzer, H., Maoz, D., Jannuzi, B.~T., \& Giveon, U.\ 2000, \apj,
    533, 631
    \bibitem[Kaspi et al.(2005)]{kaspi05} Kaspi, S., Maoz, D.,
    Netzer, H., Peterson, B.~M., Vestergaard, M., \& Jannuzi, B.~T.\
    2005, \apj, 629

   \bibitem[Kauffmann et al.(2003)]{kauffmann} Kauffmann, G., et al.\
   2003, \mnras, 346, 1055

    \bibitem[Kewley et al.(2006)]{kewley06} Kewley, L.~J., Groves,
    B., Kauffmann, G., \& Hecman, T. 2006, astro-ph/0605681

    \bibitem[Kewley et al.(2001)]{kewley01} Kewley, L.~J., Heisler,
    C.~A., Dopita, M.~A., \& Lumsden, S. 2001, \apjs, 132, 37

    \bibitem[Leighly(1999)]{leighly} Leighly, K.~M.\ 1999,
    \apjs, 125, 317

    \bibitem[Markarian et al.(1983)]{markarian} Markarian, B.~E.,
    Lipovetskii, V.~A., \& Stepanian, D.~A.\ 1983, Astrofizika, 19,
    29

    \bibitem[Martel \& Osterbrock(1994)]{martel} Martel, A.,
    \& Osterbrock, D.~E.\ 1994, \aj, 107, 1283

    \bibitem[Marziani et al.(2003)]{marziani} Marziani, P., Zamanov,
    R.~K., Sulentic, J.~W., \& Calvani, M.\ 2003, \mnras, 345, 1133

    \bibitem[Moran et al.(1996)]{moran} Moran, E.~C., Halpern, J.~P.,
    \& Helfand, D.~J.\ 1996, \apjs, 106, 341

    \bibitem[Nelson \& Whittle(1996)]{nelson96} Nelson, C.~H., \&
    Whittle, M.\ 1996, \apj, 465, 96

    \bibitem[Nelson(2000)]{nelson00} Nelson, C.~H.\ 2000, \apjl,
    544, L91

    \bibitem[Netzer(1982)]{netzer} Netzer, H.\ 1982, \mnras, 198, 589

    \bibitem[O'Donnell(1994)]{odonnell} O'Donnell, J.~E.\ 1994,
    \apj, 422, 158

    \bibitem[Onken et al.(2004)]{onken} Onken, C.~A., Ferrarese, L.,
    Merritt, D., Peterson, B.~M., Pogge, R.~W., Vestergaard, M., \&
    Wandel, A.\ 2004, \apj, 615, 645

    \bibitem[Osterbrock(1989)]{oster89} Osterbrock, D.~E.\ 1989, in
    \textit{Astrophysics of Gaseous Nebulae and Active Galactic
    Nuclei}, University Science Books, Mill Valley (CA), USA

    \bibitem[Osterbrock \& Koski(1976)]{oster76} Osterbrock, D.~E.,
    \& Koski, A.~T.\ 1976, \mnras, 176, 61P

    \bibitem[Osterbrock \& Pogge(1985)]{oster85} Osterbrock, D.~E.,
    \& Pogge, R.~W.\ 1985, \apj, 297, 166

    \bibitem[Pelat et al.(1981)]{pelat} Pelat, D., Alloin, D., \&
    Fosbury, R.~A.~E.\ 1981, \mnras, 195, 787

    \bibitem[Penston \& Perez(1984)]{penston} Penston, M.~V., \&
    Perez, E.\ 1984, \mnras, 211, 33P

    \bibitem[Peterson et al.(1985)]{peterson85} Peterson,
    B.~M., Meyers, K.~A., Carpriotti, E.~R., Foltz, C.~B., Wilkes,
    B.~J., \& Miller, H.~R.\ 1985, \apj, 292, 164

    \bibitem[Peterson(1993)]{peterson93} Peterson, B.~M.\ 1993,
    \pasp, 105, 247


    \bibitem[Rodr{\'\i}guez-Ardila et al.(2000)]{rodriguez}
    Rodr{\'\i}guez-Ardila, A., Binette, L., Pastoriza, M.~G., \&
    Donzelli, C.~J.\ 2000, \apj, 538, 581

    \bibitem[Savage \& Mathis(1979)]{savage} Savage, B.~D., \&
    Mathis, J.~S.\ 1979, \araa, 17, 73

    \bibitem[Schlegel et al.(1998)]{schlegel} Schlegel,
    D.~J., Finkbeiner, D.~P., \& Davis, M.\ 1998, \apj, 500, 525

    \bibitem[Sergeev et al.(1997)]{sergeev} Sergeev, S.~G.,
    Pronik, V.~I., Malkov, Y.~F., \& Chuvaev, K.~K.\ 1997, \aap,
    320, 405



    \bibitem[Smith et al.(2002)]{smith} Smith, J.~A., et al.\ 2002,
    \aj, 123, 2121

    \bibitem[Stepanian(2005)]{stepa05} Stepanian, J.~A.\ 2005,
    Rev. Mexicana Astron. Astrofis., 41, 155

    \bibitem[Stepanian et al.(2002)]{stepa02} Stepanian, J.~A.,
    Chavushyan, V.~H., Carrasco, L., Vald{\'e}s, J.~R., M{\'u}jica, R.~M.,
    Tovmassian, H.~M., \& Ayvazyan, V.~T.\ 2002, \aj, 124, 1283

    \bibitem[Storey \& Zeippen(2000)]{storey} Storey, P.~J., \&
    Zeippen, C.~J.\ 2000, \mnras, 312, 813

    \bibitem[Sulentic et al.(2000)]{sulentic} Sulentic, J.~W.,
    Marziani, P., \& Dultzin-Hacyan, D.\ 2000, \araa, 38, 521

    \bibitem[Tremaine et al.(2002)]{tremaine} Tremaine, S., et al.\
    2002, \apj, 574, 740

    \bibitem[Ulrich \& Horne(1996)]{ulrich} Ulrich, M.-H., \& Horne,
    K.\ 1996, \mnras, 283, 748

    \bibitem[Veilleux \& Osterbrock(1987)]{veilleux} Veilleux, S.,
    \& Osterbrock, D.~E.\ 1987, \apjs, 63, 295

    \bibitem[V{\'e}ron-Cetty et al.(2001)]{veron} V{\'e}ron-Cetty,
    M.-P., V{\'e}ron, P., \& Gon\c{c}alves, A.~C.\ 2001, \aap, 372, 730

    \bibitem[Whittet(1992)]{whittet} Whittet, D.~C.~B.\ 1992, in
    \textit{Dust in the Galactic Environment}, Tayler, R.~J. \&
    White, R.~E. eds, Institute of Physics Publishing, Bristo, UK

    \bibitem[Williams et al.(2002)]{williams} Williams, R.~J., Pogge,
    R.~W., \& Mathur, S.\ 2002, \aj, 124, 3042

    \bibitem[Wilson \& Heckman(1985)]{wilson} Wilson, A.~S., \&
    Heckman, T.~M.\ 1985, Astrophysics of Active Galaxies and
    Quasi-Stellar Objects, 39

    \bibitem[Witt et al.(1992)]{witt} Witt, A.~N., Thronson, H.~A.,
    Jr., \& Capuano, J.~M., Jr.\ 1992, \apj, 393, 611

    \bibitem[Zamanov \& Marziani(2002)]{zamanov} Zamanov, R., \&
    Marziani, P.\ 2002, \apjl, 571, L77

\end{thebibliography}
\end{document}